\documentclass[fleqn,usenatbib]{mnras}
\pdfoutput=1

\usepackage[T1]{fontenc}
\usepackage{ae,aecompl}
\usepackage{color, soul}
\usepackage{verbatim}
\usepackage{units}
\usepackage{subfig}
\usepackage{pdflscape}
\usepackage{svg}
\usepackage{bm}
\usepackage{stfloats}
\usepackage{float}
\usepackage{multicol}
\usepackage{hyperref}
\usepackage{multirow}

\usepackage{graphicx}	
\usepackage{amsmath}	
\usepackage{amssymb}	
\usepackage{epstopdf}
\usepackage[normalem]{ulem} 

\newcommand{\emr}[1] {\textcolor{red}{#1 \textsc{/EMR/}}}

\defcitealias{Lu2013}{L13}
\defcitealias{Bartko2010}{B10}
\defcitealias{Salpeter1955}{S55}
\defcitealias{Evans2022}{E22}

\newcommand\clearrow{\global\let\rowmac\relax}
\newcommand{\cmmnt}[1]{\ignorespaces}

\title[HVSs and a companion to Sgr A*]{Constraints on the Galactic Centre environment from \textit{Gaia} hypervelocity stars III: Insights on a possible companion to Sgr A*}

\author[Evans et al.]{
F. A. Evans$^{1, 2}$\thanks{E-mail: fraser.evans@utoronto.ca},
A. Rasskazov$^{3}$,
A. Remmelzwaal$^{4}$,
T. Marchetti$^{5}$,
A. Castro-Ginard$^{4}$, \newauthor
E. M. Rossi$^{4}$,
J. Bovy$^{1, 2}$ \\
$^{1}$David A. Dunlap Department of Astronomy and Astrophysics, University of Toronto, 50 St. George Street, Toronto, ON, M5S 3H4, Canada \\
$^2$Dunlap Institute for Astronomy \& Astrophysics, University of Toronto, 50 St. George Street, Toronto, ON, M5S 3H4, Canada \\
$^{2}$DAMTP, University of Cambridge, CMS, Wilberforce Road, Cambridge CB3 0WA, UK \\
$^{3}$Leiden Observatory, Leiden University, PO Box 9513, NL-2300 RA Leiden, The Netherlands\\
$^{4}$European Southern Observatory, Karl-Schwarzschild-Strasse 2, 85748 Garching bei M{\"u}nchen, Germany \\
}

\date{Accepted XXX. Received YYY; in original form ZZZ}

\pubyear{2023}

\begin{document}
\label{firstpage}
\pagerange{\pageref{firstpage}--\pageref{lastpage}}
\maketitle

\begin{abstract}

We consider a scenario in which Sgr A* is in a massive black hole binary (MBHB) with an as-of-yet undetected supermassive or intermediate-mass black hole companion. Dynamical encounters between this MBHB and single stars in its immediate vicinity would eject hypervelocity stars (HVSs) with velocities beyond the Galactic escape velocity of the Galaxy. In this work, we use existing HVS observations to constrain for the first time the existence of a companion to Sgr A*. We simulate the ejection of HVSs via the `MBHB slingshot' scenario and show that the population of HVSs detectable today depends strongly on the companion mass and the separation of the MBHB. We demonstrate that the lack of uncontroversial HVS candidates in \textit{Gaia} Data Release 3 places a firm upper limit on the mass of a possible Sgr A* companion. Within one milliparsec of Sgr A*, our results exclude a companion more massive than $1000 \, \mathrm{M_\odot}$. If Sgr A* recently merged with a companion black hole, our findings indicate that unless this companion was less massive than $500 \, \mathrm{M_\odot}$, this merger must have occurred at least $10$ Myr ago. These results complement and improve upon existing independent constraints on a companion to Sgr A* and show that large regions of its parameter space can now be ruled out.

\end{abstract}


\section{Introduction}

In the centre of our Galaxy lurks \textit{Sagittarius A*} (Sgr A*), a supermassive black hole (SMBH) with a mass of $\sim$4$\times 10^6 \, \mathrm{M_\odot}$ \citep{Ghez2008, Genzel2010, Akiyama2022}. Such SMBHs seem to be omnipresent in the nuclei of external galaxies as well \citep{Magorrian1998, Kormendy2013}, at least among galaxies above a particular mass \citep[see][]{Merritt2013}. Dynamical friction \citep{Chandrasekhar1943} in dense stellar systems conspires to bring massive bodies towards the centres of gravitational potentials. Other SMBHs delivered via major galaxy mergers or $\sim10^2 - 10^4 \, \mathrm{M_\odot}$ intermediate mass black holes (IMBHs) formed originally in stellar clusters \citep{Miller2002, Portegies2002} or directly in a SMBH accretion disc \citep{Goodman2004, McKernan2012} can thereby form a bound \textit{massive black hole binary} (MBHB) consisting of either an SMBH-SMBH pair \citep[e.g.][]{Begelman1980, Volonteri2003, DiMatteo2005} or an SMBH-IMBH pair \citep{Levin2005, Portegies2006, ArcaSedda2018}. MBHBs have received considerable attention in recent years owing to the fact that colliding MBHBs are loud gravitational wave sources \citep{Peters1964} in a frequency range accessible to the upcoming Laser Interferometer Space Antenna \citep[LISA;][]{AmaroSeoane2017}.

There exists the possibility that Sgr A* has a SMBH or IMBH companion which, owing to the fact that the Galactic Centre (GC) makes for a challenging observational environment \citep[see][for a review]{Schodel2014}, has not yet been detected. The possible parameter space for such a companion has been steadily reduced over the past two decades from astrometric observations of Sgr A* itself \citep{Hansen2003, Reid2004, Reid2020}, orbital modelling of the S-star cluster \citep{Gualandris2009} and of the S-star cluster star S2 in particular \citep{Gualandris2010, Naoz2020, GRAVITY2020}. Taken together, current constraints allow for a companion of at most a few hundred $\mathrm{M_\odot}$ just within or outside the orbit of S2 (c.f. \citealt{GRAVITY2020}, Appendix D).

An independent method of constraining a possible companion to Sgr A* is by considering its direct interactions with stars that approach it. Dynamical interactions between single stars and MBHBs tend to `fling' the stars out at roughly the orbital velocity of the less massive member of the MBHB, draining energy from the system and hardening the MBHB \citep{Quinlan1996, Zier2001}. Ejection velocities therefore increase as the MBHB hardens, to the point where the hard MBHB is able to eject so-called `hypervelocity stars' (HVSs) with velocities in excess of the Galactic escape speed \citep{Yu2003}. It should be noted that a companion to Sgr A* is not required to produce HVSs in the first place -- the term HVS was coined by \citet{Hills1988}, who first theorized that the disruption of a stellar binary by a single SMBH could eject one member of the binary with an extreme velocity \citep[see also][]{Yu2003, Kenyon2008, Sari2010, Kobayashi2012, Rossi2014, Zhang2013, Generozov2020, Generozov2021}. While this so-called Hills mechanism remains the most promising scenario for the ejection of fast stars from the centre of the Galaxy, the `MBHB slingshot' mechanism described above is a well-researched alternative \citep{Yu2003, Levin2003, Baumgardt2006, Sesana2006, Sesana2007, Lockmann2008, Marchetti2018, Rasskazov2019, Darbha2019, Zheng2021, MastrobuonoBattisti2023}.  

After the predictions of \citet{Hills1988}, nearly two decades would elapse until the first detection of a promising HVS candidate by \citet{Brown2005}, who discovered a B-dwarf in the outer Galactic stellar halo with a heliocentric radial velocity of $853 \pm 12 \, \mathrm{km \ s^{-1}}$. Following this and other serendipitous HVS discoveries \citep{Edelmann2005, Hirsch2005}, other candidate stars potentially unbound to the Galaxy have trickled in courtesy of targeted surveys \citep{Brown2006, Brown2009, Brown2012, Brown2014}, follow-up observations of previously identified stars \citep[e.g][]{Heber2008,Tillich2009,Irrgang2010,Irrgang2019} and queries of large Galactic surveys \citep[e.g.][]{Palladino2014, Zhong2014, Huang2017, Koposov2020}. See \citet{Brown2015rev} for a review of these objects. While the current roster of proposed HVS candidates stands at $\sim$two dozen, only S5-HVS1 \citep{Koposov2020} can be uncontroversially associated with an origin in the Galactic Centre.

Our understanding of the structure and kinematics of the Milky Way as a whole has been revolutionized by the European Space Agency satellite \textit{Gaia} \citep{Gaia2016}. In its third and most recent data release \citep[DR3;][]{Gaia2022DR3}, \textit{Gaia} provides five-parameter astrometry (parallax, position, proper motion) for $\sim$1.5 billion Galactic sources \citep{Gaia2020EDR3} and validated heliocentric radial velocities for $\sim$34 million sources \citep{Katz2022, Sartoretti2022}. While \textit{Gaia} observations have been invaluable in the identification of new potential HVS candidates \citep{Bromley2018, Shen2018, Hattori2018, Du2019, Li2018, Luna2019, Huang2021, Li2021, Li2022, Igoshev2022, Prudil2022}, currently missing from all data releases of \textit{Gaia} radial velocity catalogues are high-confidence HVS candidates, that is, candidates with i) precise astrometric measurements (relative parallax error $<$20\%), ii) a velocity which suggests the star is unbound to the Galaxy, and iii) a trajectory which is consistent with an origin in the Galactic Centre \citep{Marchetti2019, Marchetti2021, Marchetti2022}.

This presents a tantalizing opportunity. If Sgr A* has a hidden companion, HVSs ejected via the MBHB slingshot mechanism may be detectable by \textit{Gaia}. Given that the selection functions of both the \textit{Gaia} source catalogue and radial velocity atalogue are relatively well-modelled \citep[][Castro-Ginard et al. submitted]{Everall2022, Sartoretti2022, CantatGaudin2022}, it is possible to determine with reasonable confidence which stars should appear in the catalogue and which stars should not.  An absence of confident HVS candidates in \textit{Gaia} DR3 can therefore rule out areas of the MBHB parameter space (namely the hidden companion's mass and separation from Sgr A*) which predict an abundance of MBHB slingshot-ejected HVSs.

This work is a companion to \citet{Evans2022a, Evans2022b} and \citet{Marchetti2022}, in which we exploited a similar possibility, focusing on the Hills mehanism. In \citet{Evans2022a} we used the lack of main sequence HVSs in \textit{Gaia} Early Data Release 3 \citep{Gaia2020EDR3} to impose an upper limit on the HVS ejection rate. We expanded upon this in \citet{Evans2022b}, showing that considering the lack of \textit{evolved} HVS candidates improves constraints. By considering as well the existence of S5-HVS1 \citep{Koposov2020}, strengthened constraints further and additionally constrained the shape of the initial mass function (IMF) among HVS progenitor binaries in the GC. In \citet{Marchetti2022} we mined the radial velocity catalogue of the then-newly released \textit{Gaia} DR3 to search for new high-confidence HVS candidates. While no new HVS candidates were unearthed with precise astrometry, with a Galactocentric velocity in excess of $700 \ \mathrm{km \ s^{-1}}$ and with a trajectory pointing away from the GC, we showed how the lack of HVS candidates in this data release improved constraints even further. In those works we considered the Hills mechanism as the \textit{only} mechanism which ejects HVSs. In this work we follow a similar philosophy considering the MBHB slingshot mechanism -- by exploring how different present-day MBHB binary configurations would result in HVS populations of different size, we show that some configurations are incompatible with an absence of detected HVSs in the \textit{Gaia} DR3 radial velocity sample.

This paper is organized as follows. In Sec. \ref{sec:methods} we outline on a step-by-step basis our model for generating mock populations of HVSs ejected via the MBHB slingshot mechanism. In Sec. \ref{sec:results} we present our analyses of these populations -- we show how the population of HVSs in \textit{Gaia} depends on the characteristics of the MBHB binary and how the \textit{absence} of high-confidence HVS candidates in \textit{Gaia} DR3 constrains the mass of an as-of-yet unseen companion to Sgr A* and its separation from it. We discuss these results in Sec. \ref{sec:discussion} before offering a summary and conclusions in Sec. \ref{sec:conclusions}.

\section{Massive black hole binary Slingshot Ejection Model} \label{sec:methods}

In this section we describe how we model the MBHB orbital evolution, how we generate ejected HVSs, how we propagate these HVSs through the Galaxy and how we obtain mock observations of them to determine which would be detectable by \textit{Gaia}. Overall our procedure resembles the approach of \citet{Marchetti2018}, who also explore \textit{Gaia}-detectable HVSs ejected from an MBHB in the GC. We expand upon their modelling by investigating a larger set of MBHB configurations and by using updated prescriptions for propagating HVSs through the Galaxy, performing mock photometry, and modelling the \textit{Gaia} selection function. The code we use to implement the model we describe in this section is included as the \texttt{MBHB} module in the publicly available \texttt{PYTHON} package \texttt{speedystar}\footnote{\url{ https://github.com/fraserevans/speedystar}}.


\subsection{Orbital decay of the MBHB}

An MBHB embedded in a collisionless, fixed stellar background loses orbital energy 
via i) the ejection of stars via the MBHB slingshot mechanism, and ii) the emission of gravitational waves. We model the hardening of the MBHB during phase i)  following \citet{Quinlan1996}:

\begin{equation} \label{eq:dadt_HVS}
    \frac{da}{dt}\bigg|_{\rm HVS} = - \frac{G \rho H}{\sigma}a^{2} \, \text{,}
\end{equation}
where $G$ is the gravitational constant, $\rho$ and $\sigma$ are the mass density and one-dimensional velocity dispersion of the stellar background, respectively, assumed here to be $\rho=7\times 10^{4} \, \mathrm{M_\odot \ pc^{-3}}$ \citep{Schodel2007, Feldmeier2014, Schodel2014} and $\sigma = 100 \, \mathrm{km \ s^{-1}}$ \citep{Figer2003, Schodel2009, Feldmeier2014, Do2020} and $H$ is a dimensionless hardening rate:

\begin{equation}
    H(\sigma, \rho, a) = \frac{\sigma}{G \rho}\frac{d}{dt}\left( \frac{1}{a} \right) \, \text{.}
\end{equation}
As noted by \citet{Quinlan1996}, $H$ is approximately constant when the binary separation is below the `hardening separation' $a_h$: 

\begin{equation} \label{eq:ah}
    a_{h}\equiv \frac{GM_{\rm c}}{4\sigma^2} \; \text{,}
\end{equation} 
where $M_{\rm c}$ is the mass of the less-massive companion in the MBHB. We approximate the orbital decay of the MBHB due to gravitational wave emission following \citet{Peters1964}:

\begin{equation} \label{eq:dadt_GW}
    \frac{da}{dt}\bigg|_{\rm GW} = - \frac{64}{5}G^3 c^5 \frac{M_{\rm Sgr A^*}  M_{\rm c}  M_{\rm total}}{a^3} \, \text{,}
\end{equation}
where $c$ is the speed of light, $M_{\rm Sgr A^*}$ is the mass of Sgr A*, taken here as $4\times10^{6} \, \mathrm{M_\odot}$ \citep{Eisenhauer2005, Ghez2008}, and $M_{\rm total} \equiv M_{\rm Sgr A^*} + M_{\rm c}$ is the total mass of the MBHB. The evolution of the MBHB semi-major axis with time is then

\begin{equation}
    a(t) = a_0 + \int_{t_0}^{t} \left(\frac{da}{dt}\bigg|_{\rm HVS} + \frac{da}{dt}\bigg|_{\rm GW}\right)dt \, \text{,}
\end{equation}
where at an initial time $t_0$ the binary starts at separation $a=a_0$. 

Following as well from \citet{Quinlan1996}, the total mass $\Delta M_{\rm ej}$ in stars ejected by the MBHB as it shrinks from separation $a$ to $a-\Delta a$ is modelled using the dimensionless mass ejection rate $J$: 

\begin{equation} \label{eq:dMej}
    \Delta M_{\rm ej} = J M_{\rm total} \Delta \mathrm{ln}(1/a) \, \text{.}
\end{equation}

Note that this ejection rate above assumes the MBHB orbital decay is driven \textit{entirely} by HVS ejections. To account for the contribution of GW emission to the binary hardening, we must include a correction to Eq. \ref{eq:dMej}:

\begin{equation} \label{eq:dMejcorr}
    \Delta M_{\rm ej} = \frac{\Delta t_{\rm HVS}^{-1}}{\Delta t_{\rm total}^{-1}} J M_{\rm total} \Delta\text{ln}(1/a) \; \text{,}
\end{equation}
where $\Delta t_{\rm total}$ is the total time required for the MBHB to shrink from separation $a$ to $a-\Delta a$ and $\Delta t_{\rm HVS}>\Delta t_{\rm total}$ is the time required to shrink from $a$ to $a-\Delta a$ if the decay is entirely by HVS ejections, calculated by integrating and inverting Eq. \ref{eq:dadt_HVS}. 

\begin{figure*}
    \centering
    \includegraphics[width=\columnwidth]{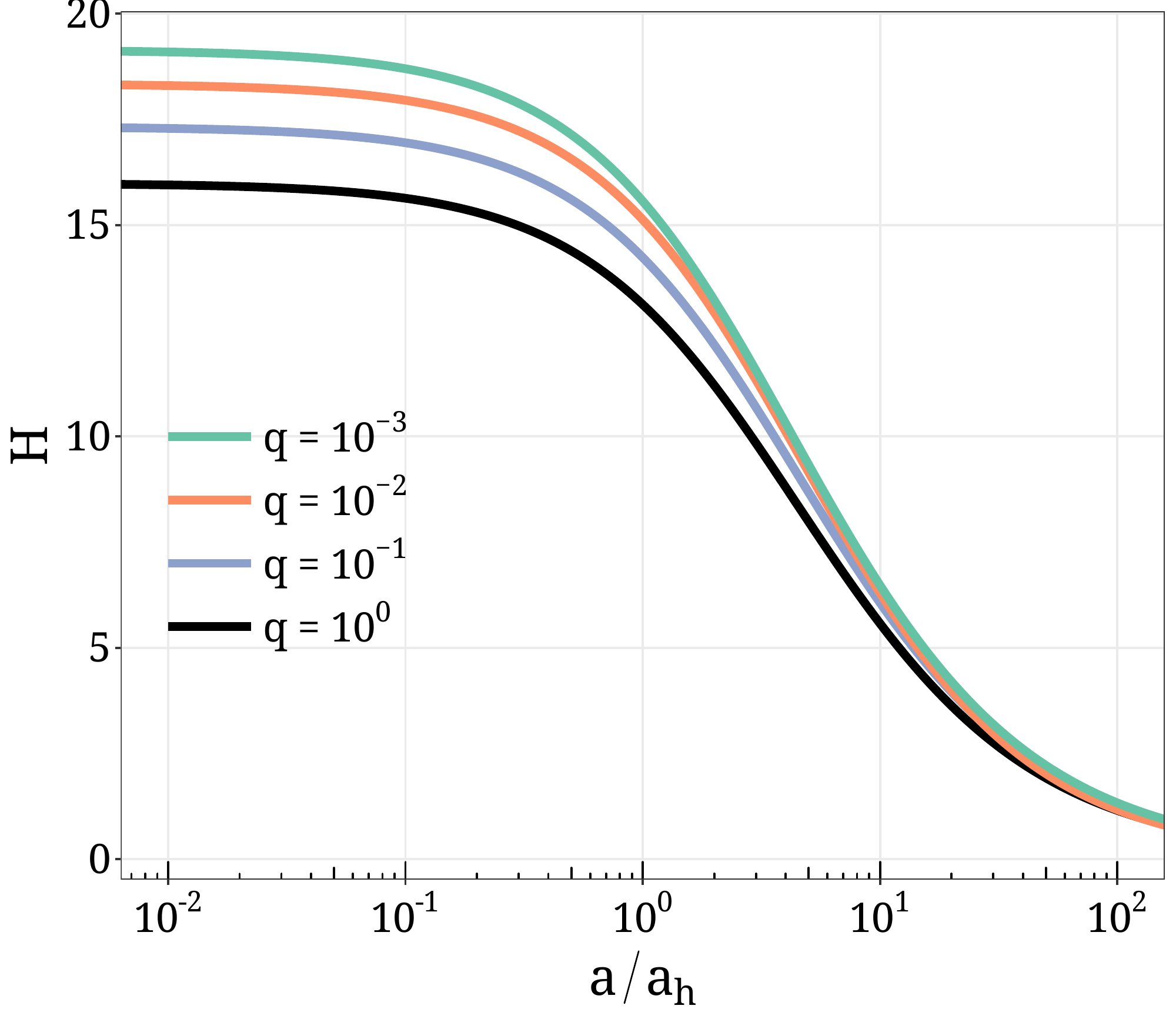}
    \includegraphics[width=\columnwidth]{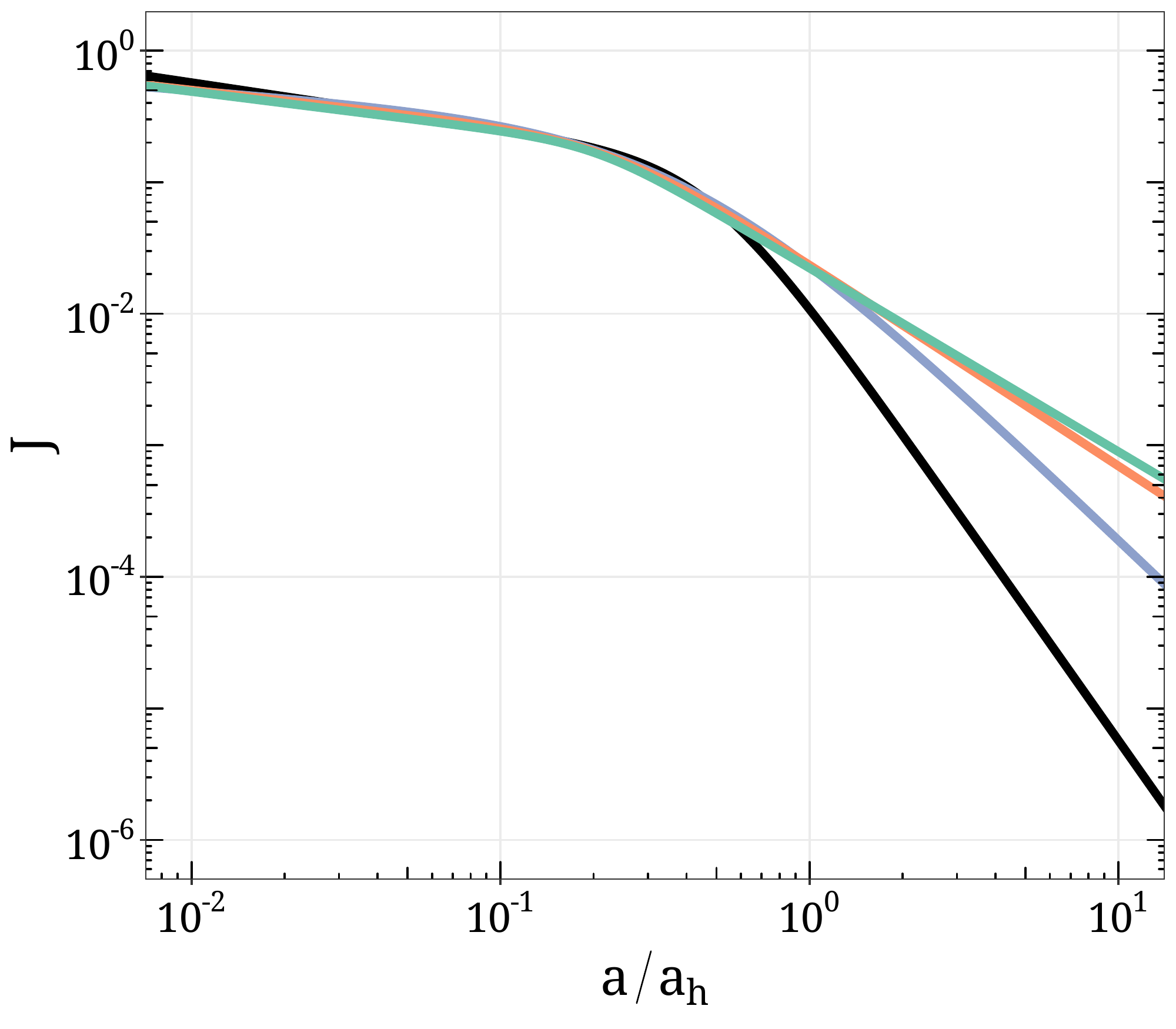}
    \caption{The dependence of the dimensionless hardening rate $H$ (left) and mass ejection rate $J$ (right) on the MBHB hardness, assuming a circular MBHB orbit. MBHB separations are scaled to the hardening separation (Eq. \ref{eq:ah}) Relationships for MBHBs with differing mass ratios ($q$) are shown with different line colours.}
    \label{fig:HJ}
\end{figure*}

Over a grid of MBHB mass ratios $q$ in the range $10^{-4}\leq q \leq 10^0$, we determine the dimensionless hardening rate $H$ and the mass ejection rate $J$ by perform scattering experiments. The methodology of these experiments closely follows \citet{Rasskazov2019} and we refer the reader to that work for more details. Stars, assumed to be massless, approach a MBHB of separation $a$ from infinity with a velocity $v$, impact parameter $b$, on directions randomized with respect to the binary phase and orbital plane. $v$ is drawn from a Maxwellian distribution in the range [$3\times10^3v_0\sqrt{q/(1+q)}, 30 v_0\sqrt{q/(1+q)}$] \citep{Sesana2006}, where
\begin{equation} \label{eq:v0}
v_0 \equiv \sqrt{\frac{G(M_{\rm c} + M_{\rm MBH})}{a}}    
\end{equation}
is the orbital velocity of the MBHB. $b^2$ is drawn uniformly such that pericenter distances are constrained to the range $[0,5a]$. We assume the scattering interactions never increase the eccentricity of the binary -- its orbit remains circular throughout (see discussion on this point in Sec. \ref{sec:discussion}). Each star is scattered off the binary and the simulation ends when i) the star reaches a distance of 50$a$ from the MBHB with a positive total energy, or ii) the timescale for the scattering interaction exceeds 10 Gyr, or iii) when the star spends longer than $1.6\times10^4$ binary orbital periods within a distance of 50$a$ from the MBHB. For each MBHB mass ratio $q$ we run a total of four million such simulations using the ARCHAIN \citep{Mikkola2008}  algorithm, specifically developed to simulate small-$N$ systems.
    
The results of these simulations are shown in Fig. \ref{fig:HJ}, where we plot how $H$ and $J$ depend on the MBHB separation for various mass ratios. In our calculation of $J$, a star counts as `ejected' if and only if its velocity at a large distance from the MBHB is larger than $5.5\sigma=550 \; \mathrm{km \ s^{-1}}$, which is approximately the escape velocity from the bulge if it is modelled as a single isothermal sphere profile. In general $H$ at fixed $a/a_{\rm h}$ increases with decreasing $q$ though the relationship is not strictly monotonic. Note as well that more massive binaries still merge more quickly since $a_{\rm h}$ decreases linearly with $q$. In agreement with \citet{Sesana2006}, we find that each can be approximated as 
    
\begin{align} \label{eq:HJfits}
H(a) &= A_H ( 1 + a/a_{0, H} )^{\gamma_H} \; \text{,} \\
J(a) &= A_J (a/a_{0, J})^{\alpha_J} ( 1 + (a/a_{0, J})^{\beta_J})^{\gamma_J} \; \text{,}
 \end{align}   
where $A$, $\alpha$, $\beta$, $\gamma$ are fitting parameters. In Table \ref{tab:fits} we share best-fit parameters for a selection of mass ratios.

\begin{table*}
\centering 
\caption{Best fit parameters for the hardening rate $H$ and stellar mass ejection rate $J$ (see Eq. \ref{eq:HJfits}).}
\label{tab:fits}                    
\begin{tabular}{l|c|c|c|c|c|c|c|c|}        
\hline               
mass ratio & $A_H$ & $a_{0, H}/a_h$ & $\gamma_H$ & $A_J$ & $a_{0, J}/a_h$ & $\alpha_J$ & $\beta_J$ & $\gamma_J$ \\    
\hline
1.0 & 14.58 & 1.98 & -0.62 & 0.21 & 	0.14 & -0.36 & 11.17 & -0.20 \\
0.3	& 15.98 & 3.42 & -0.77 & 0.15 & 	0.48 & -0.35 & 2.77 & -1.08 \\
0.1	& 17.33 & 3.43 & -0.77 & 0.26 & 0.41 & 	-0.18 & 1.32 & -1.57 \\
0.03 & 18.34 & 3.86 & -0.84 & 0.22 & 0.24 & -0.27 & 2.37 & -0.54 \\
0.01 & 19.14 & 3.24 & -0.77 & 0.20 & 0.21 & -0.29 & 3.75 & -0.29 \\
0.003 & 18.27 & 4.00 & -0.81 & 0.21 & 0.20 & -0.30 & 4.31 & -0.26 \\
0.001 & 17.29 & 3.29 & -0.75 & 0.22 & 0.21 & -0.30 & 3.93 & -0.30 \\
0.0003 & 17.36 & 2.92 & -0.73 & 0.22 & 0.22 & -0.29 & 3.46 & -0.36 \\
\hline                                   
\end{tabular}
\end{table*}

Another result of our scattering experiments, echoing the results of \citet{Rasskazov2019}, is that the high-velocity tail of the ejection velocity distribution ($v_{\rm ej}\geq 400 \, \mathrm{km \ s^{-1}}$) is well-fit by a broken power-law distribution of the form

\begin{equation} \label{eq:vej}
    \frac{\mathrm{dN}}{\mathrm{dlog}v_{\rm ej}} =
    \begin{cases}
        C_{\rm 1}, & \text{if} \ v_{\rm ej} < v_{\rm break} \\ 
        C_{\rm 2}v_{\rm ej}^{-3.1}, & \text{if} \ v_{\rm ej} \geq v_{\rm break} 
    \end{cases}
\end{equation}
where $C_1$ and $C_2$ are constants, $v_{\rm break} = 1.2v_0\sqrt{2q}/(1+q)$ and $v_0$ is the circular velocity of the MBHB (Eq. \ref{eq:v0}). When assigning initial velocities to ejected stars, we draw velocities at random from this distribution in the range [5.5$\sigma$, 5$v_0$].

\begin{figure}
    \centering
    \includegraphics[width=\columnwidth]{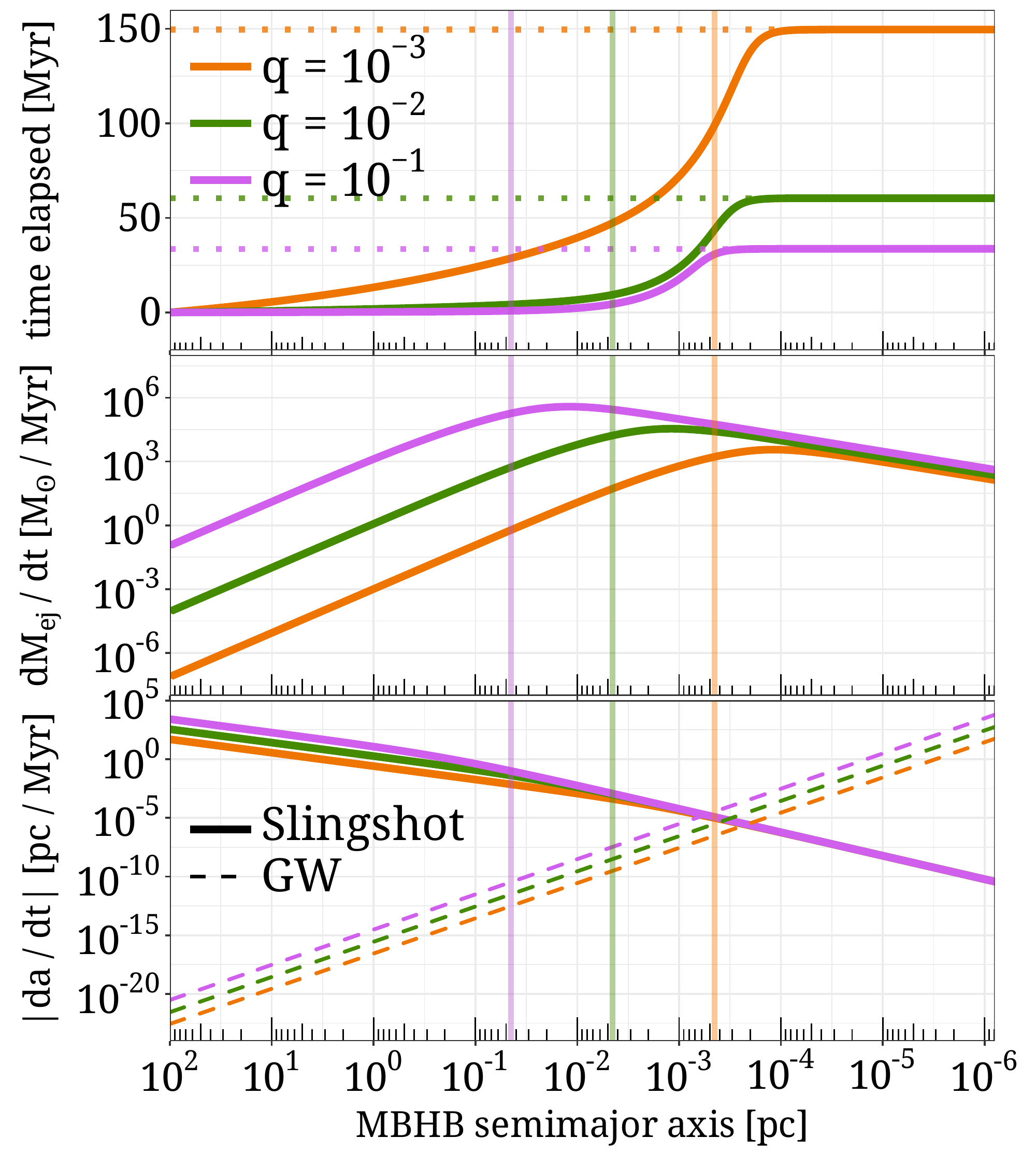}
    \caption{The MBHB orbital decay and rate of  mass in ejected stars as a function of binary semi-major axis, for three values of the mass ratio $q$. The coloured vertical lines show the hardening separation (Eq. \ref{eq:ah}) for the corresponding mass ratio. The primary MBH mass in each scenario is assumed to be $4\times10^{6} \, \mathrm{M_\odot}$ and each starts at $t=0$ at a separation $a_0 = 100 \, \mathrm{pc}$. The top panel shows how the separation shrinks with time, with the dotted horizontal lines showing when each binary merges. The middle panel shows how the stellar mass ejection rate evolves throughout the inspiral. The bottom panel shows how the decay rate changes with decreasing separation, with the contributions from HVS ejections and from GW emissions plotted separately. }
    \label{fig:orbit}
\end{figure}

In Fig. \ref{fig:orbit} we summarize our modelling of the MBHB orbit, showing the orbital decay and ejected stellar mass for MBHBs with $q=10^{-1}$, $q=10^{-2}$ and $q=10^{-3}$. Each binary has an initial separation of $a_0 = 100 \, \mathrm{pc}$ at $t=0$, and for the purposes of this study it is sufficient to say the MBHB has `merged' when its separation reaches $10^{-6} \, \mathrm{pc}$, as after this the separation will drop to zero within a year. The top panel shows the elapsed time required for the MBHB to shrink to a separation $a$. The total time to merge is quite sensitive to the companion mass -- the $q=10^{-1}$, $q=10^{-2}$ and $q=10^{-3}$ binaries merge within 34, 60 and 120 Myr, respectively. The vertical lines show the hardening separations $a_{\rm h}$ for the corresponding MBHB. The MBHB reaches $a=a_{\rm h}$ at earlier stages of the inspiral for progressively more massive companions of Sgr A*. Middle panel shows that the HVS mass ejection rate profile peaks shortly after $a=a_{\rm h}$. For small $q$ this peak occurs after $da/dt|_{\rm HVS} = da/dt|_{\rm GW}$ (see bottom panel) and as $q$ increases this peak occurs while the slingshot mechanism is still the dominant hardening mechanism. The total ejected HVS mass also depends strongly and nonlinearly on the MBHB companion mass -- a binary with a mass ratio of $10^{-2}$ ejects 330 times more HVS mass in total than a binary with a mass ratio of $10^{-4}$, but only eight times more mass than a binary with a mass ratio of $10^{-3}$.

\subsection{Generating the HVS sample}

The previous subsection described how the mass ejected in HVSs by a MBHB is related to the characteristics and evolution of the binary. Here we describe how we apply this to generate individual mock HVSs. Our approach is as follows.

We assume a MBHB composed of Sgr A* and a companion of mass $M_{\rm c}$ started initially at separation $a_0 = 100 \, \mathrm{pc}$\footnote{This is a very large separation, chosen mostly to ensure we use an initial separation which captures all HVS ejections. In practice, we find that only stars ejected when the MBHB is at a separation of $\lesssim$ 0.1 pc are detectable in any \textit{Gaia} data release.} at time $t_0$ and is now at separation $a_{\rm now}$. We create a grid of 1000 semi-major axes from $a_{\rm now}$ to $a_0$ uniformly spaced in log-space.
We use Eq. \ref{eq:dMejcorr} to determine the mass $\Delta M_{\mathrm{ej}, i}$ ejected while the binary was in the $i$'th separation bin, $i$ = 1, 2, ..., 1000. The corresponding \textit{number} of ejected HVSs is

\begin{equation}
    \Delta N_i = \frac{\Delta M_{\mathrm{ej}, i}}{\int_{M_{\rm min}}^{M_{\rm max}} Mf(M)dM} \; \text{,}
\end{equation}
where $f(M)$ is the assumed IMF, defined between minimum and maximum stellar masses $M_{\rm min}$ and $M_{\rm max}$. We adopt a single power-law IMF with slope $\kappa$, i.e. $f(m) \propto m^{-\kappa}$ and minimum and maximum masses of $0.1 \, \mathrm{M_\odot}$ and $100 \, \mathrm{M_\odot}$, respectively.

If the MBHB is currently at separation $a_{\rm now}$, the flight time of all stars ejected while the binary was in the $i$'th separation bin have the same flight time $t_{\rm flight, \textit{i}}$:

\begin{equation}
    t_{\rm flight, \textit{i}} = \int_{a_i}^{a_{\rm now}} \left(\frac{da}{dt}\bigg|_{\rm HVS} + \frac{da}{dt}\bigg|_{\rm GW}\right)^{-1}da \, \text{,}
\end{equation}
where $a_i$ is the midpoint of the $i$'th separation bin. We assume a star is equally likely to be ejected at any point in its lifetime, therefore we say the age of each star at ejection $t_{\rm age, ej}$ is a random fraction of its maximum lifetime $t_{\rm life}$;
\begin{equation}
    t_{\rm age, ej} = \epsilon \cdot t_{\rm life} \; ,
\end{equation}
where $\epsilon$ is a random number uniformly distributed in [0,1]. $t_{\rm life}$ for each ejected star is determined using the single stellar evolution \citep[SSE;][]{Hurley2000} algorithms within the \texttt{AMUSE}\footnote{\url{https://amuse.readthedocs.io/en/latest/index.html}}environment \citep{Portegies2009,Portegies2013,Pelupessy2013,Portegies2018}, taking the start of the asymptotic giant branch phase as the `end' of a star's life. We cap $t_{\rm life}$ to a maximum of 13.8 Gyr to ensure the star is not older than the Universe.

Today, the age of each star is 
\begin{equation} \label{eq:tage}
    t_{\rm age} = t_{\rm age,ej} + t_{\rm flight} \; .
\end{equation}
We remove stars for whom $t_{\rm age}>t_{\rm life}$, i.e. stars which were ejected as main sequence or evolved stars but are stellar remnants in the present day.

After drawing an ejection velocity for each surviving star following Eq. \ref{eq:vej}, stars are initialized on a sphere $3 \, \mathrm{pc}$ in radius centred on Sgr A* with a velocity pointing radially away from the GC. We assume stars are ejected isotropically -- we comment further on this assumption in Sec. \ref{sec:discussion}. 

\subsection{Orbital integration and mock photometry} \label{sec:methods:potential}

After initializing our mock HVSs, our scheme for propagating them forward in time through the Galactic potential and obtaining synthetic photometry of them remains essentially unchanged from the method described in \citet{Evans2022b}. We refer the reader to that work for more detailed explanations and briefly summarize the approaches here.

We propagate each star forward in time for its flight time assuming the Galactic potentials of \citet{McMillan2017}, who use a Monte Carlo Markov Chain (MCMC) method to fit a many-component potential to various kinematic data. For each realization, we draw a potential at random from the \citet{McMillan2017} MC chain (P. McMillan, private communication). All ejected stars are integrated through this potential with the \texttt{PYTHON} package \texttt{GALPY}\footnote{\url{https://github.com/jobovy/galpy}} \citep{Bovy2015} using a fifth-order Dormand-Prince integrator \citep{Dormand1980} and a timestep of $0.1 \, \mathrm{Myr}$.  

We estimate the visual dust extinction at each star's distance and sky position using the \textsc{combined15} dust map of \citep{Bovy2016}\footnote{\url{https://github.com/jobovy/mwdust}}, itself a combination of the Galactic dust maps of \citet{Drimmel2003, Marshall2006} and \citet{Green2015}. We then determine mock apparent magnitudes for each mock HVS in the photometric bands using the MESA Isochrone and Stellar Tracks, or \texttt{MIST} \citep{Dotter2016, Choi2016} models\footnote{\url{https://waps.cfa.harvard.edu/MIST/}}. With each star's luminosity, surface gravity and effective temperature (determined with \texttt{AMUSE} depending on each star's mass, age and metallicity), as well as its visual extinction, we interpolate the \texttt{MIST} bolometric correction tables to determine each star's apparent magnitude in the Johnsons-Cousins $V$ and $I_{\rm c}$ bands \citep{Bessell1990} and \textit{Gaia} $G$ and $G_{\rm RP}$, bands\footnote{\url{see https://www.cosmos.esa.int/web/gaia/edr3-passbands}} \citep{Riello2021}. We estimate each star's magnitude in the \textit{Gaia} $G_{\rm RVS}$ band from its $V$, $I_{\rm c}$ and $G$-band magnitudes using fits from \citet{Jordi2010}.

\subsection{HVSs identifiable by \textit{Gaia}}
\label{sec:methods:Gaia}

With mock magnitudes and stellar parameters for our synthetic HVS populations, we now determine which HVS candidates should be detectable in the different \textit{Gaia} data releases with measured radial velocities. While we use an updated selection function, this approach is similar in philosophy to \citet{Evans2022b}.

For \textit{Gaia} DR3, we start by discarding all HVSs with effective temperatures outside the range $3500 \, \mathrm{K} \leq  T_{\rm eff} \leq 6900 \, \mathrm{K}$, since the \textit{Gaia} spectroscopic pipeline does not assign validated radial velocities to stars outside this temperature range \citep{Katz2022, Sartoretti2022}. Next, we use the selection functions made available by the GaiaUnlimited\footnote{\url{https://github.com/gaia-unlimited/gaiaunlimited}} project \citep{CantatGaudin2022, CastroGinard2023} to identify stars which would be detectable in the \textit{Gaia} DR3 radial velocity catalogue. In short, the DR3 empirical selection function is calibrated against the Dark Energy Camera Plane Survey Data Release 1 \citep{Schlafly2018}, and querying it yields the probability $p_{\rm source}$ that a star at a given sky position with a given \textit{Gaia G}-band magnitude would appear in the DR3 source catalogue, i.e. it would have at least a measured position, magnitude and colour. Since it compares to a deeper survey, We note that this approach only characterizes the faint-end selection function of \textit{Gaia}. At the bright end, saturation effects lead to incompleteness for sources brighter than $G\lesssim3$ \citep[see][]{Fabricius2021, Gaia2021}. We remove all mock HVSs brighter than $G=3$, however we note that it is extraordinarily rare for our model to produce HVSs this bright. Likewise, querying the spectroscopic selection function yields the probability $p_{\rm vrad}$ that a star in the DR3 source catalogue at a given sky position, $G$-band magnitude and $G-G_{\rm RP}$ colour is in the radial velocity catalogue as well. We manually set $p_{\rm vrad}$ to zero for mock HVSs in sky position/magnitude/colour bins entirely unpopulated in the DR3 radial velocity catalogue, since in the GaiaUnlimited statistical model $p_{\rm vrad}$ will be non-zero but small and highly uncertain. The total probability $p$ that a mock HVS would appear in the \textit{Gaia} DR3 radial velocity subsample is then the product of $p_{\rm source}$ and $p_{\rm vrad}$.  For each mock HVS in our sample we draw a random number $0 < \epsilon < 1$ and designate the HVS as DR3-detectable if $\epsilon < p $.


After deciding which mock HVSs would have radial velocities in DR3, we determine each one's 5x5 astrometric covariance matrix (position, parallax and proper motion uncertainties and correlations amongst them) by querying the \textit{Gaia} DR3 astrometric spread function of \citet{Everall2021cogiv}\footnote{see \url{https://github.com/gaiaverse/scanninglaw}}, which computes uncertainties based on the sky position and $G$-band magnitude of the source. Finally, we determine the DR3 radial velocity uncertainty for each star using the \texttt{PYTHON} package \texttt{PyGaia}\footnote{\url{https://github.com/agabrown/PyGaia}}.



We also determine which stars will be detectable in the fourth\textit{Gaia} data release, DR4. Radial velocity measurements in this survey would be available for all mock HVSs cooler than $6900 \, \mathrm{K}$ down to the $G_{\rm RVS}=16.2$ mag faint-end magnitude limit of the \textit{Gaia} radial velocity spectrometer \citep{Cropper2018, Katz2019}. For hotter HVSs, radial velocities would be available for stars brighter than $G_{\rm RVS}=14$. We estimate DR4 astrometric errors using \citet{Everall2021cogiv} DR3 astrometric spread function, reducing the errors according to the predicted \textit{Gaia} performance\footnote{\url{https://www.cosmos.esa.int/web/gaia/science-performance}, see also \citet{Brown2019}.}. 

To be labelled as a `detectable' high-confidence HVS in a particular \textit{Gaia} DR3 or DR4, an ejected star must:

\begin{itemize}
    \item be brighter than the faint-end apparent magnitude limit of the radial velocity catalogue of the data release.
    \item have an effective temperature range within the bounds imposed by each data release (see above). 
    \item have a relative parallax uncertainty below 20\%. For larger uncertainties, estimating distances (and therefore total velocities) becomes problematic \citep[see][]{BailerJones2015}.
    \item have a total velocity in the Galactocentric rest frame in excess of $700 \, \mathrm{km \ s^{-1}}$. We sample over the astrometric and radial velocity uncertainties of each mock star and compute its total velocity assuming a distance from the Sun to the GC of $8.122$ kpc \citep{GRAVITY2018}, a height of the Sun above the Galactic disc of $20.8$ pc \citep{Bennett2019} and a rest-frame velocity of the Sun of $\textbf{v}_\odot \equiv [U_\odot, V_\odot, W_\odot] = [12.9, 245.6, 7.78] \, \mathrm{km \ s^{-1}}$ \citep{Reid2004, Drimmel2018}. Upon sampling, in at least 80\% of realizations the total velocity of the star must be above $700 \, \mathrm{km \ s^{-1}}$.

\end{itemize}    

These above cuts resemble those used to search for HVS candidates in \textit{Gaia} Data Release 2 \citep{Marchetti2019}, Early Data Release 3 \citep{Marchetti2021} and DR3 \citep{Marchetti2022}. We remark that \citet{Marchetti2019}, \citet{Marchetti2021} and \citet{Evans2022a, Evans2022b} selected HVSs by searching for stars likely to be moving faster than the Galactic escape speed at their position. In \citet{Marchetti2022} and in this work, however, we focus instead on stars moving faster than $700 \, \mathrm{km \ s^{-1}}$. Determining the boundedness of a star requires assuming a Galactic potential, whereas a flat velocity cut instead allows agnosticism towards the potential. This cut at $700 \, \mathrm{km \ s^{-1}}$ is conservative -- in reasonable models of the Milky Way potential the escape velocity is <$700 \, \mathrm{km \ s^{-1}}$ everywhere except perhaps the within the innermost kpc of the Galaxy in heavier models \citep[e.g.][]{McMillan2017}.

For brevity, in the remainder of this work we use the terms `\textit{Gaia} DR3/DR4' to refer exclusively to the radial velocity subsamples, and by the term `HVS' we refer only to those stars which satisfy the criteria listed above.

\section{Results} \label{sec:results}

\subsection{An existing companion to Sgr A*}

\begin{figure*}
    \centering
    \includegraphics[width=0.65\columnwidth]{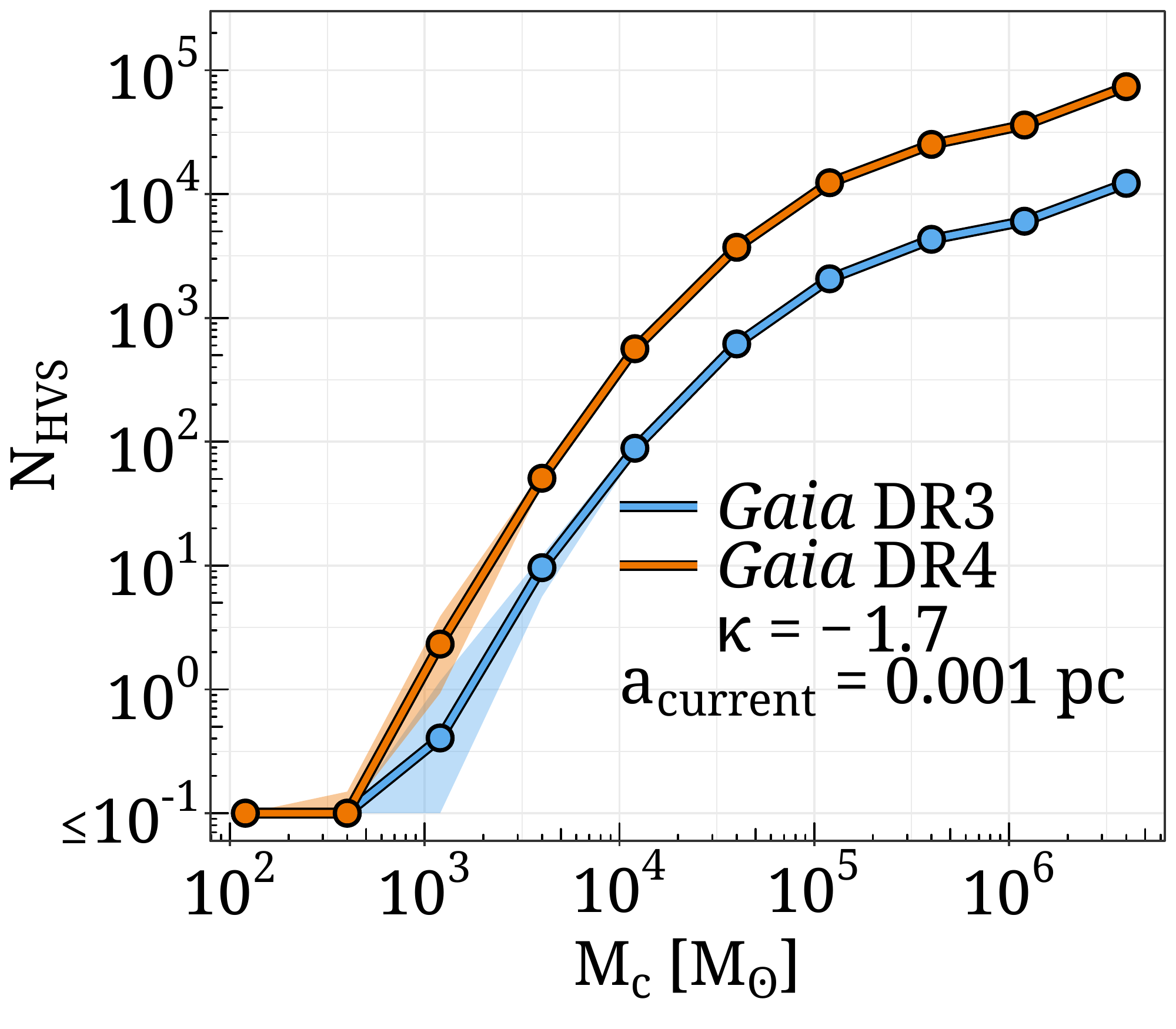}
    \includegraphics[width=0.65\columnwidth]{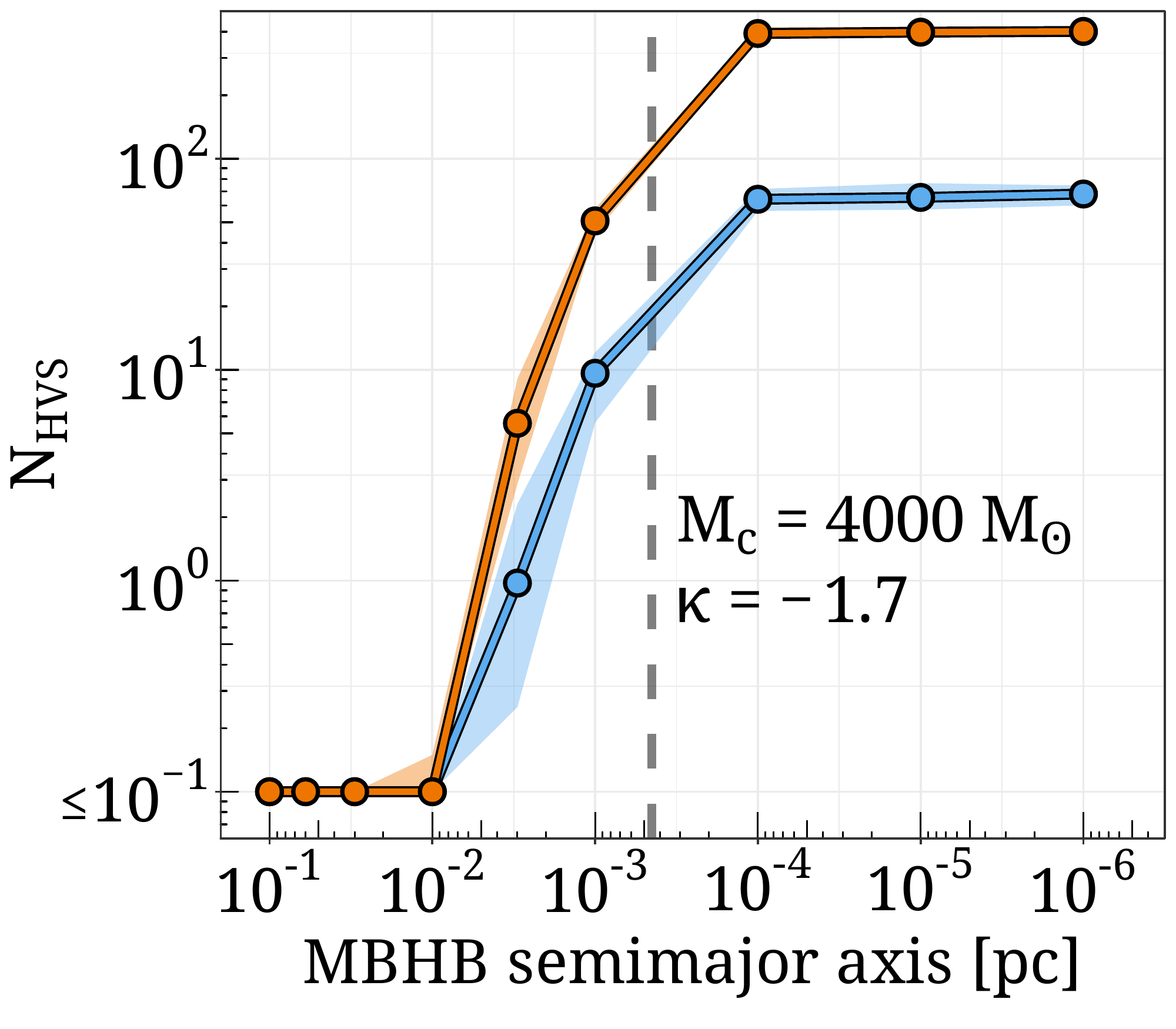}    
    \includegraphics[width=0.65\columnwidth]{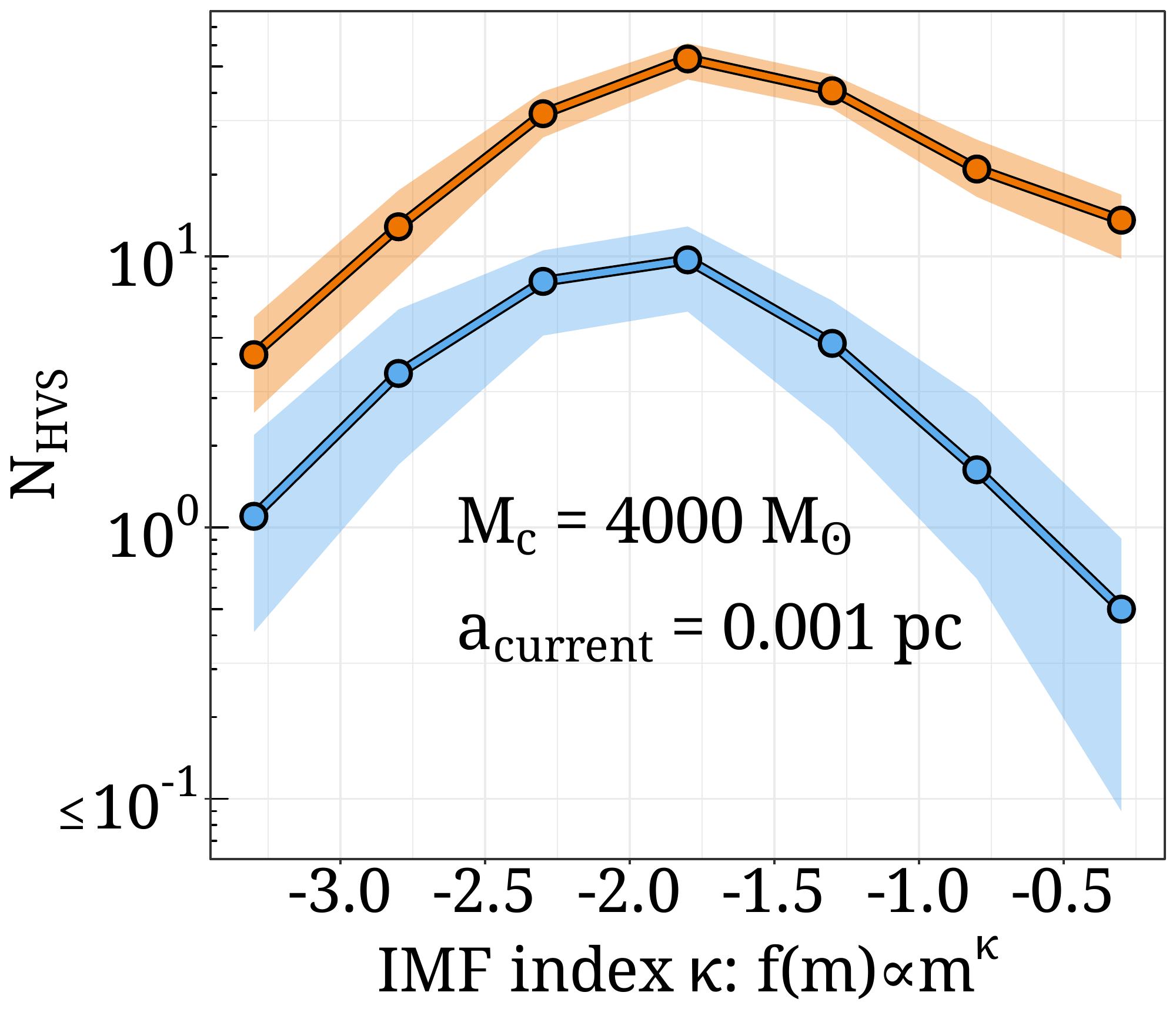}
    
    \caption{The population of high-confidence HVSs (see Sec. \ref{sec:methods:Gaia}) expected to appear in \textit{Gaia} DR3 and DR4. Panels show how N$_{\rm HVS}$ depends on the mass of the companion (left), the current separation $a_{\rm current}$ between Sgr A* and its companion (middle) and the slope $\kappa$ of and initial mass function among HVS progenitors (right). Parameters are fixed to their fiducial values when not being varied. Shaded regions span the 16th to 84th quantiles over 40 iterations. The vertical dashed line in the middle panel shows the hardening separation for this binary.}

    \label{fig:Nums}
\end{figure*}

\begin{figure}
    \centering
    \includegraphics[width=\columnwidth]{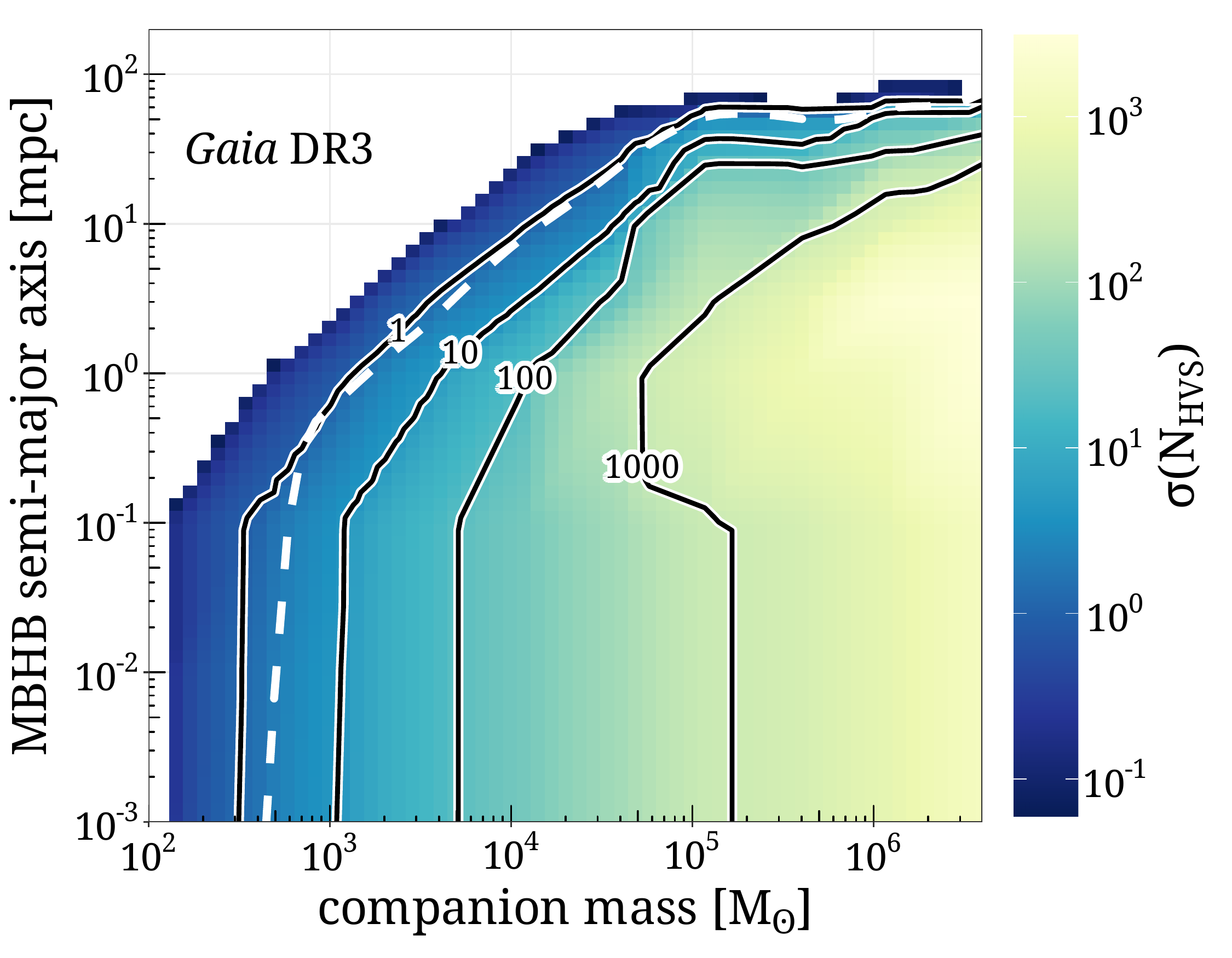}
    \caption{Contour lines show how the population of HVSs detectable in Gaia DR3 depends on the Sgr A* companion mass $M_{\rm c}$ and the current separation $a_{\rm current}$ between Sgr A* and its companion, averaged over 40 realizations and smoothed over the grid. The colourbar shows how the 1$\sigma$ scatter of N$_{\rm HVS}$. The white dashed line shows where the 1$\sigma$ lower bound of $N_{\rm HVS}$ reaches one -- the parameter space \textit{above} this line is consistent with zero HVS detections in \textit{Gaia} DR3.}
    \label{fig:McaDR3}
\end{figure}

Having outlined our model for generating mock HVS populations from the MBHB slingshot mechanism, we start in Fig. \ref{fig:Nums} by illustrating how the HVS population detectable in \textit{Gaia} DR3 and DR4 depends on our model assumptions. In the left panel we show how the \textit{Gaia} HVS population depends on the IMBH companion mass when the MBHB separation is fixed at $1 \, \mathrm{mpc}$ and the IMF of HVS progenitors has a power-law slope of -1.7 \citep{Lu2013}. Below a companion mass of $\simeq800 \, \mathrm{M_\odot}$, we expect $\leq 1$ HVS in both DR3 and DR4. This expected DR3 population can rise to several tens of thousands if Sgr A* is in a near-equal mass MBHB. The steep dependence on M$_{\rm c}$ is fairly intuitive -- as the companion mass increases, the total ejected stellar mass increases and the ejection velocity distribution shifts towards higher velocities (Fig. \ref{fig:orbit}, Eq. \ref{eq:vej}). A strong dependence on the current separation $a_{\rm current}$ of the MBHB (middle panel) is digestible as well -- a smaller current separation means the MBHB has ejected more HVSs in the recent past, and HVS ejection velocities are larger since the MBHB circular velocity is larger. $N_{\rm HVS}$ plateaus for separations smaller than the binary hardening separation (vertical line) because the binary merges shortly thereafter (see Fig. \ref{fig:orbit}). For an HVS IMF slope of -1.7 and a companion mass of $4000 \, \mathrm{M_\odot}$, >1 HVS is expected in \textit{Gaia} DR3 as long as the MBHB separation is less than a few milliparsecs. This number should reach $\sim$60 in DR3 or several hundred in DR4 if the MBHB is just about to merge within the next decade\footnote{A small point of possible confusion -- for simplicity, we assume the MBHB separation does not shrink throughout the \textit{Gaia} mission lifetime. Given typical typical MBHB coalescence times (see Fig. \ref{fig:orbit}), this assumption is valid. The analysis presented in this work assumes a scenario in which \textit{Gaia} DR3 and DR4 are `snapshots' of the Milky Way at a time when the MBHB separation is $a_{\rm current}$, rather than observations collected over the span of 34 or 66 months, respectively.}.

In the right panel of Fig. \ref{fig:Nums} we show the dependence of N$_{\rm HVS}$ on the IMF power law slope $\kappa$. The turnover at $\kappa\simeq -1.7$ can be explained by two competing factors. Firstly, as $\kappa$ increases, ejected stars are on average more massive, and therefore more luminous and more likely to be brighter than \textit{Gaia}'s faint-end magnitude limit of $\sim14$ for the DR3 radial velocity sample. Secondly, however, the total number of ejected HVSs \textit{decreases} as $\kappa$ increases, since the ejected stellar mass is locked up in fewer, more massive stars (see Eq. \ref{eq:dMejcorr}). The turnover represents the point where the latter effect overcomes the former. 

Since the impact of $\kappa$ on N$_{\rm HVS}$ in DR3 is comparatively weak (particularly in the vicinity of $\kappa\approx -1.7$), for the remainder of this work we explore only constraints on $a_{\rm current}$ and $M_{\rm c}$. We run simulations over a $M_{\rm c}/a_{\rm current}$ grid, marginalizing over $\kappa$ by sampling it at random in each iteration from a Gaussian distribution centred on $\kappa=-1.7$ with a standard deviation of 0.2 \citep{Lu2013}. In Fig. \ref{fig:McaDR3} we show the results of this simulation suite. The contours show lines of constant $N_{\rm HVS}$ throughout $M_{\rm c}-a_{\rm current}$ space and the colourbar indicates the 1$\sigma$ scatter. For our fiducial choices of $M_{\rm c} = 4000 \, \mathrm{M_\odot}$ and $a_{\rm current} = 1 $ mpc, we determine that $8\pm4$ high-confidence HVSs should have been uncovered in the radial velocity catalogue of \textit{Gaia} DR3. Following from Fig. \ref{fig:Nums}, the expected DR3 HVS population ranges from <1 (for low-mass companions at large separations) to $\sim$thousands (for a near-equal-mass, hardened MBHB). The relative scatter is $\sigma(N_{\rm HVS})/N_{\rm HVS}\simeq0.2$ for $N_{\rm HVS}\gtrsim 100$, rising to $\simeq0.5$ and $\simeq1$ if $\sim$tens or only a few HVS were uncovered, respectively. For $M_{\rm c} \lesssim 10^5 \, \mathrm{M_\odot}$ $N_{\rm HVS}$ increases monotonically with decreasing $a_{\rm current}$. For larger companion masses, however, $N_{\rm HVS}$ decreases for $a_{\rm current}$ less than $\sim$ a few $\times 10^{-1}$ mpc since the MBHB slingshot mass ejection rate peaks and begins to decline before GW emission takes over and the binary quickly merges (see Fig. \ref{fig:orbit} middle panel). The dashed white line indicates where the 1$\sigma$ lower limit of $N_{\rm HVS}$ is equal to one, i.e. where one HVS \textit{at least} should be in the survey. Since no such HVSs have been detected in \textit{Gaia} DR3, MBHB binaries below this line can be excluded. A companion to Sgr A* cannot exist within 1 mpc (2000 AU) of the GC unless it is quite low-mass ($M_{\rm c}\lesssim$2000 $\mathrm{M_\odot}$). A near-equal mass MBHB remains possible only for separations $\gtrsim 0.05 \mathrm{pc}$.

\begin{figure}
    \centering
    \includegraphics[width=\columnwidth]{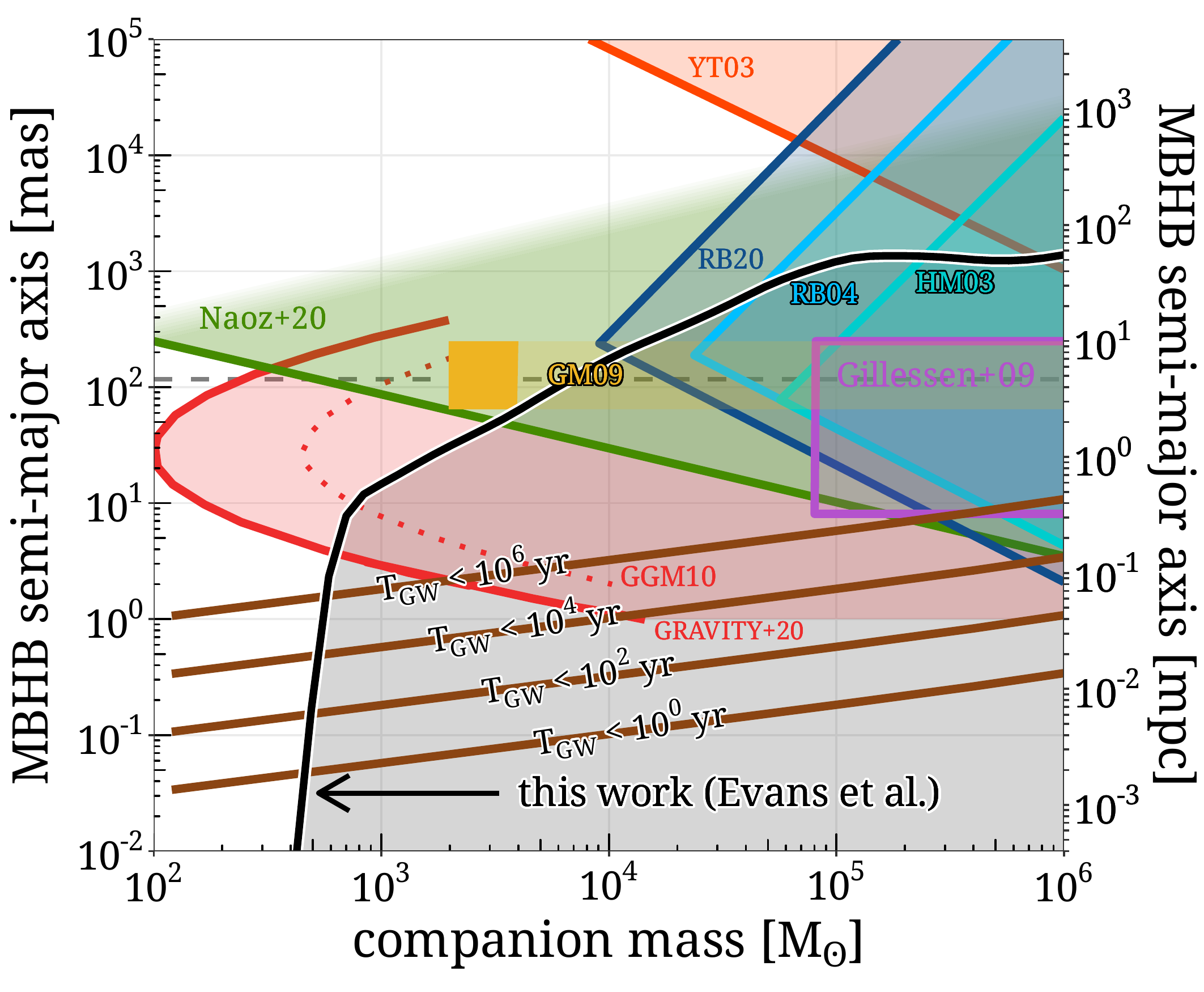}
    \caption{Constraints on the mass of a black hole companion to Sgr A* and its separation from Sgr A* in angle (left axis) and physical space (right axis). Adapted from \citet{GRAVITY2020}, adapted in turn from \citet{Gualandris2009}. The lack of confident HVS candidates in \textit{Gaia} DR3 excludes the region below the black line. The dashed horizontal line highlights the semimajor axis of S2 at 125 mas \citep[$\approx 4.7 \, \mathrm{mpc}$][]{GRAVITY2020}. \citet{Yu2003} previously excluded the orange region by remarking that the barycentre of a presumed MBHB in the GC cannot be significantly displaced from Sgr A*, otherwise the cusp of stellar density in the GC would not be coincident with Sgr A*. The cyan, blue and dark blue regions are excluded from astrometric observations of Sgr A* \citep{Hansen2003IMBH, Reid2004, Reid2020}. The violet region is excluded by the orbit of S2 \citep{Gillessen2009}. The green region is excluded by \citet{Naoz2020} by requiring that the orbit of S2 is stable against perturbations by a Sgr A* companion (a fading colour denotes weakening constraints). The solid gold region is excluded by \citet{Gualandris2009} using the orbital eccentricity distribution of the S-star cluster. Companion masses to the right of the dotted red line are excluded by \citet{Gualandris2010} from the orbit of S2. Following the 2018 pericentric passage of S2, these constraints were improved by \citet{GRAVITY2020}. The diagonal dotted lines show lines of constant gravitational wave inspiral time for the MBHB. }
    \label{fig:CompanionConstraints}
\end{figure}

Prior works have constrained a companion to Sgr A*. In Fig. \ref{fig:CompanionConstraints} we place our constraints in context with these. In all, large regions of the parameter space we exclude in this work have previously been excluded, particularly cases where Sgr A* has a fairly massive companion just inside or outside the orbit of S2. These configurations, if true would result in an astrometric `wobble' of Sgr A* \citep{Hansen2003IMBH, Reid2004, Reid2020},  would impact the orbit of S2 \citep{Gillessen2009}, its stability against perturbation \citep{Gualandris2010, Naoz2020, GRAVITY2020} and the stability of the S-star cluster in general \citep{Gualandris2009}. HVS observations are an independent tool to measure the Galactic Centre and the constraints they impose on a possible MBHB in the GC reinforce and somewhat sharpen these prior constraints. 

Of the parameter space that has \textit{not} already been excluded by previous works, we mainly rule out companions separated from Sgr A* by less than a milliparsec. Such configurations, depending on the mass of the companion, could only persist for a short amount of time between gravitational wave emission drives the MBHB to coalescence. We annotate Fig. \ref{fig:CompanionConstraints} with lines of constant gravitational wave coalescence times. Some prior works have disregarded configurations below these lines outright, as it would be fortuitous if, in the present day, Sgr A* were about to merge with a companion in the very near future. Regardless, a lack of HVS candidates in \textit{Gaia} DR3 observations exclude such fortunate scenarios and demonstrates the advantage of using HVS observations alongside traditional probes. Combining prior constraints with constraints determined in this work, the allowed configurations of a MBHB in the GC include i) a $\lesssim 5\times10^4 \, \mathrm{M\odot}$ companion at a separation greater than a few tens of milliparsecs from Sgr A*, ii) a $M = 100-200 \, \mathrm{M_\odot}$ companion just within or outside the orbit of S2 (970 AU, or 4.7 mpc), or iii) a $M\lesssim 500 \, \mathrm{M_\odot}$ companion at a separation of less than a few hundred microparsecs from Sgr A*.

\subsection{A \textit{former} companion to Sgr A*}

\begin{figure*}
    \centering
    \includegraphics[width=0.92\columnwidth]{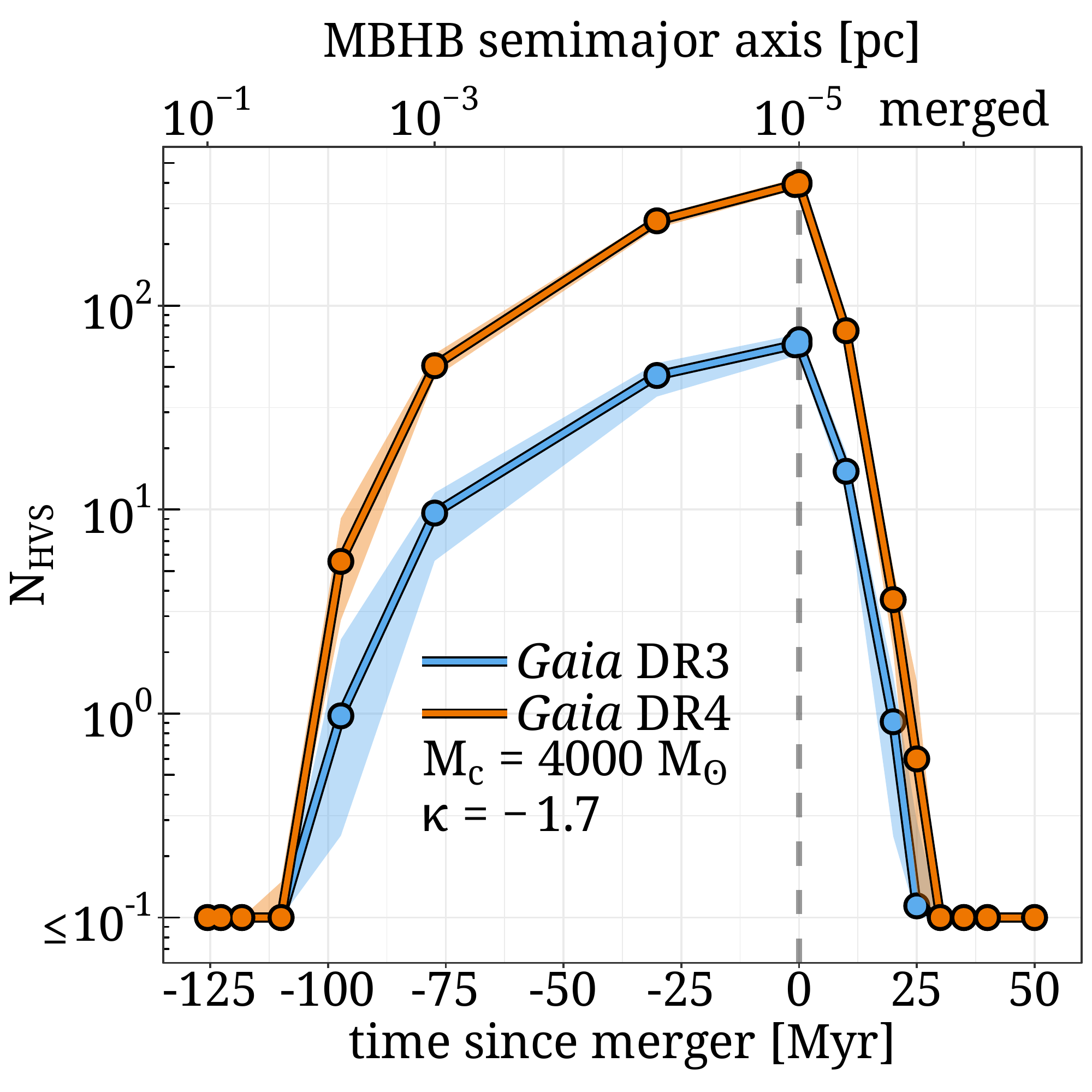}
    \includegraphics[width=1.05\columnwidth]{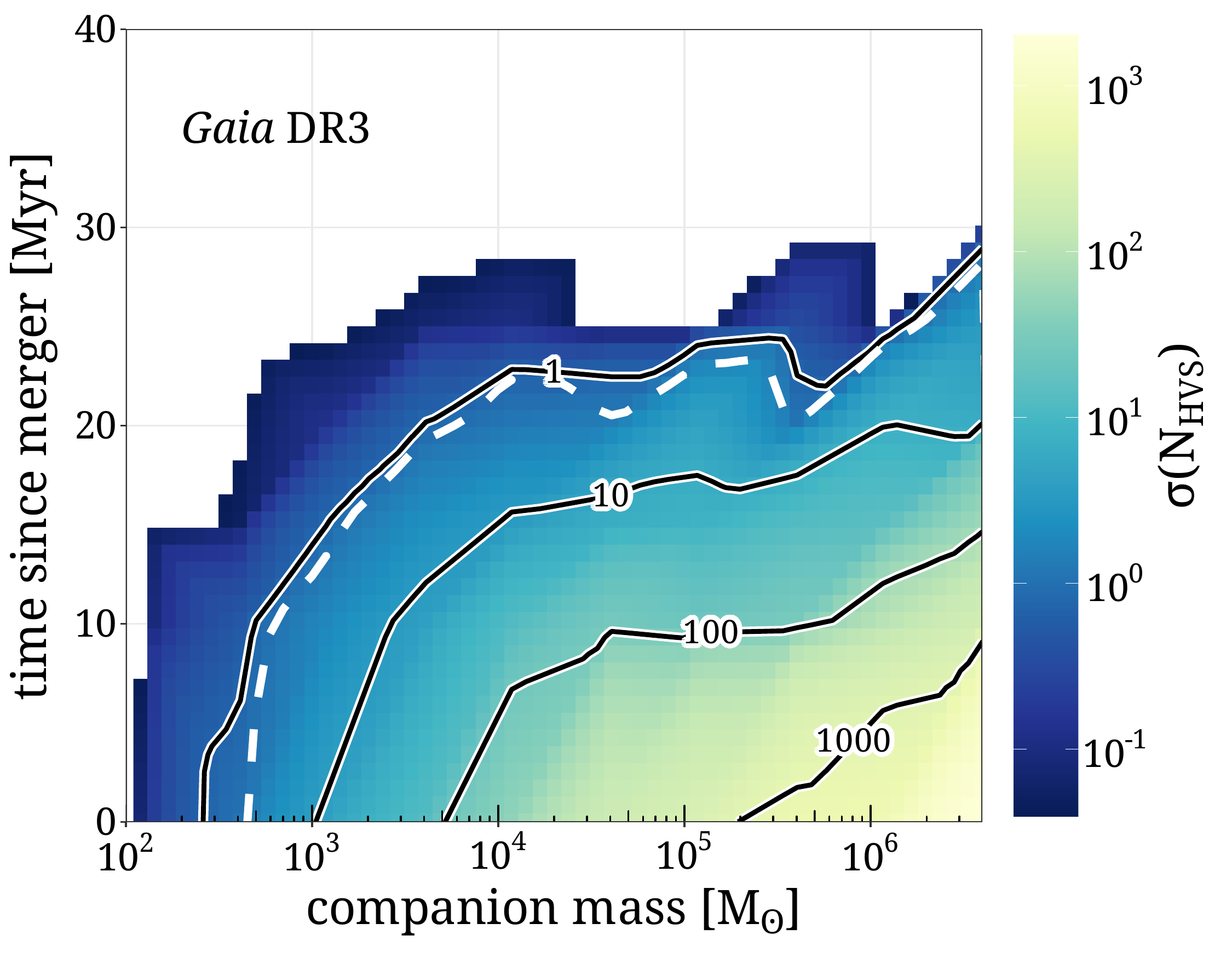}
    \caption{\textit{Left}: The dependence of $N_{\rm HVS}$ on the time $t_{\rm since}$ since the MBHB has merged with a companion. A negative $t_{\rm since}$ indicates that the MBHB has not yet merged -- when this is the case, the top horizontal axis indicates the MBHB separation. Shaded regions span the 16th to 84th quantiles over 50 iterations. \textit{Right:} contour lines show how the population HVSs detectable in Gaia DR3 depends on the former Sgr A* companion mass $M_{\rm c}$ and time elapsed since the merger $t_{\rm since}$, averaged over 20 realizations and smoothed over the grid. The colourbar shows how the 1$\sigma$ scatter of N$_{\rm HVS}$. The dashed line shows where the 1$\sigma$ lower bound of $N_{\rm HVS}$ equals one.}
    \label{fig:MctDR3}
\end{figure*}

As the MBHB spirals in, the HVS mass ejection rate peaks at a small separation \citep[Fig. \ref{fig:orbit}, see also][]{Gualandris2005, Baumgardt2006, Levin2006}. Post-merger, this final `gasp' of HVSs propagates outwards through the Galaxy. Over the next $\sim$tens of Myr, these HVSs can still be detected from Earth. This means HVS observations can probe not only an existing companion to Sgr A*, but a former companion as well. We illustrate this in Fig. \ref{fig:MctDR3}. In the left panel we show how $N_{\rm HVS}$ depends on the time $t_{\rm since}$ since the MBHB in the GC merged, holding $M_{\rm c}$ and the IMF index $\kappa$ fixed. A negative $t_{\rm since}$ indicates the binary has not yet merged. $N_{\rm HVS}$ is maximized just before the MBHB coalesces. Otherwise, from the HVS population size alone one cannot determine  whether the MBHB exists in the present day or whether it already merged in the recent past. Given a substantial population of HVS candidates, the distribution of flight times could discriminate between these two scenarios (see Sec. \ref{sec:discussion}).

In the right panel of Fig. \ref{fig:MctDR3} we show how $N_{\rm HVS}$ depends on both $M_{\rm c}$ and $t_{\rm since}$ when we sample over $\kappa$. The dashed line indicates where the 1$\sigma$ lower limit of $N_{\rm HVS}$ reaches one. Any configuration below this line can be ruled out, since at least one HVS ejected before the MBHB merged should still be detectable in \textit{Gaia} DR3. If Sgr A* ever had a companion more massive than $1000 \, \mathrm{M_\odot}$, it must have merged with Sgr A* more than 12 Myr ago. This lower limit on $t_{\rm since}$ increases with increasing companion mass. To our knowledge this is the first direct observational constraint on the specific merger history of Sgr A* within the last $\sim$tens of Myr. If Sgr A* merged with a companion more than 30 Myr ago, we cannot offer constraints on how massive that companion could have been. similarly, if Sgr A* recently accreted a companion less massive than $\sim$200 $\mathrm{M_\odot}$, current HVS observations cannot constrain how recently this merger occurred. 

\subsection{Prospects for DR4} 

\begin{figure*}
    \centering
    \includegraphics[width=0.9\columnwidth]{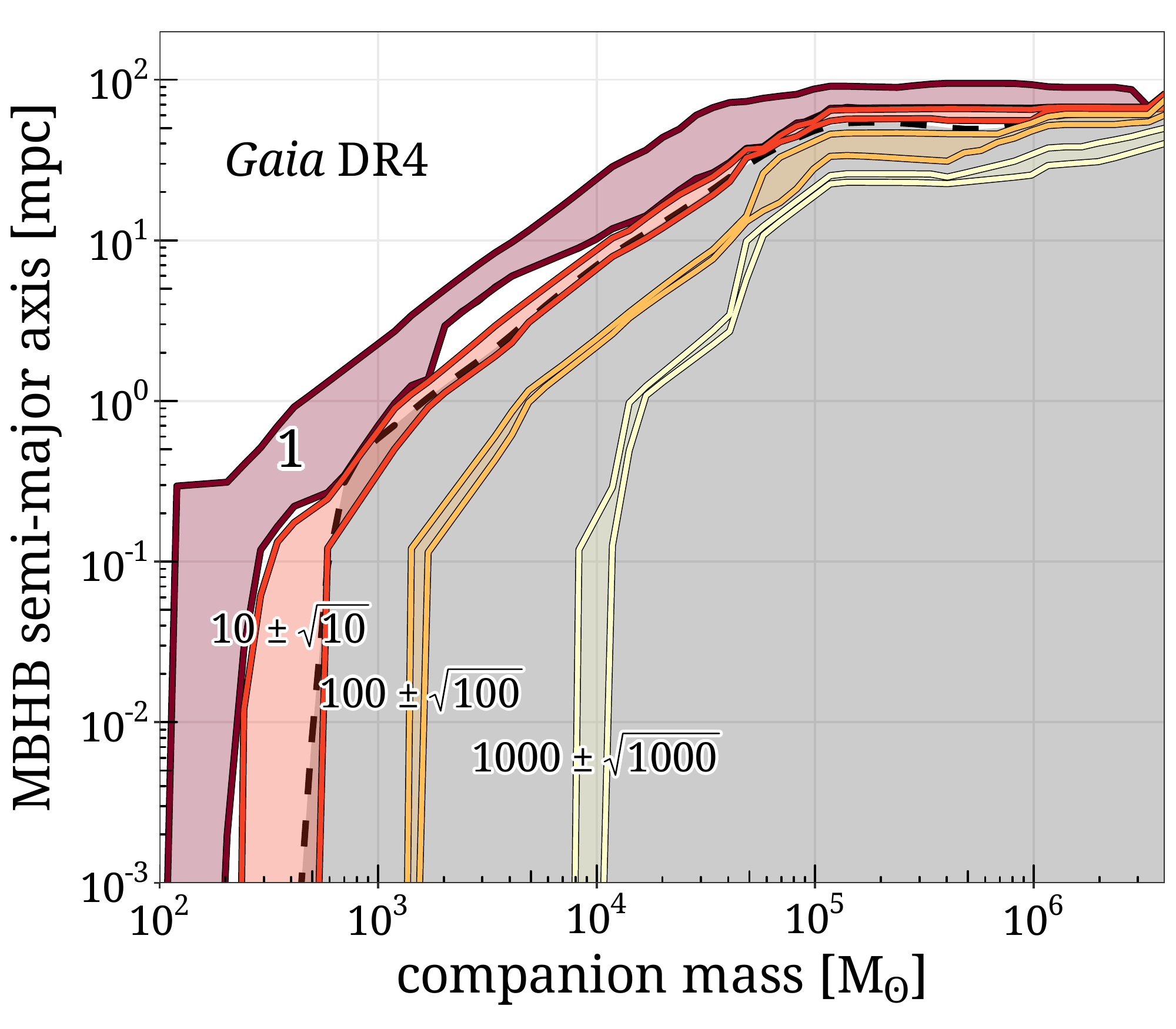}
    \includegraphics[width=0.9\columnwidth]{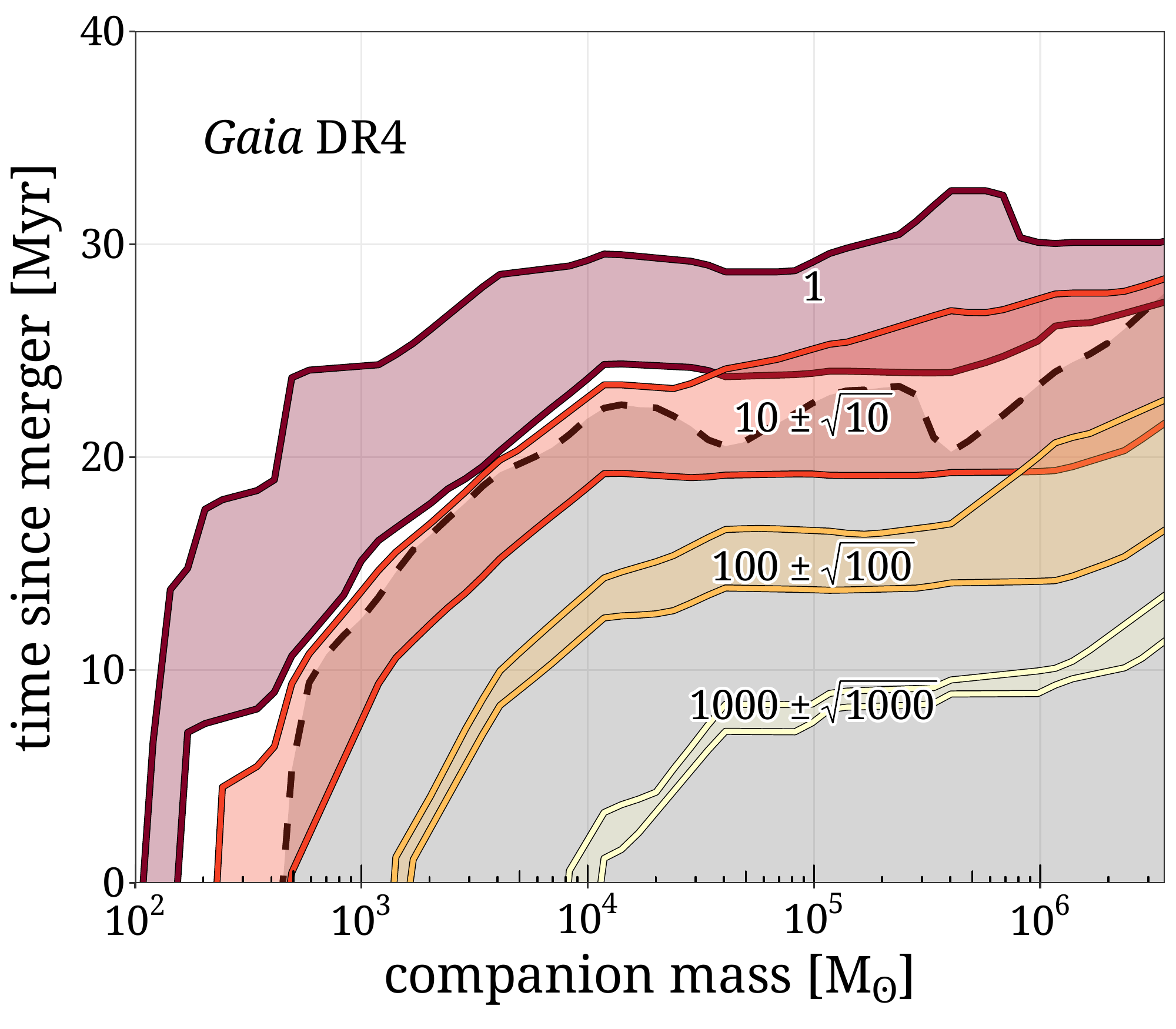}
    \caption{The coloured bands show configurations of $M_{\rm c}$ and $a_{\rm current}$ (left) or $M_{\rm c}$ and $t_{\rm since}$ (right) space consistent within 1$\sigma$ with finding the labelled number of HVSs in \textit{Gaia} DR4. The shaded grey regions are already excluded by the lack of HVSs in \textit{Gaia} DR3.}
    \label{fig:NumsDR4}
\end{figure*}

Having established constraints on a possible existing or former Sgr A* companion from \textit{Gaia} DR3 HVS observations (or lack thereof), we can project forward and explore how these constraints may update with the improved astrometric precision and deeper faint-end magnitude limit of \textit{Gaia} DR4, expected $\sim$2026. In Fig. \ref{fig:NumsDR4} we show how constraints on the presumed (existing or former) Sgr A* companion mass and either its current separation from Sgr A* (if the MBHB still exists) or the time since coalescence (if the MBHB has already merged) will change depending on the number  of high-confidence HVS uncovered in \textit{Gaia} DR4. The coloured bands show the parameter space allowed if $1$, $10\pm \sqrt{10}$, $100\pm 10$ or $1000 \pm \sqrt{1000}$ HVSs are found in DR4. Given the lack of HVSs in \textit{Gaia} DR3, our simulations suggest we should expect no more than 9, 17, or 40 HVSs in DR4 if the companion mass is $10^3$, $10^4$, or $10^5 \, \mathrm{M_\odot}$ respectively. If zero high-confidence HVSs are discovered in \textit{Gaia} DR4, regions of parameter space below the lower edge of the `1' band will be excluded. Otherwise, since $N_{\rm HVS}$ depends so intimately on $M_{\rm c}$ and $a_{\rm current}$/$t_{\rm since}$, strict (albeit degenerate) constraints can be placed on these parameters if a non-zero number of HVSs are uncovered in DR4. 


\section{Discussion}
\label{sec:discussion}

\subsection{Contribution from the Hills mechanism}

The results of the previous section assume that \textit{all} detectable HVSs are ejected via the MBHB slingshot mechanism. This is not necessarily (nor is it expected to be) true, as multiple possible HVS ejection mechanisms exist and given a sizeable HVS population it may be difficult to disentangle which was ejected via which mechanism. In practice, the ejection of HVSs from the GC may to due to a myriad of mechanisms, each occurring in tandem and coupling to each other. These other possible avenues include the dynamical encounters between single stars and a `swarm' of stellar-mass black holes in the GC \citep{OLeary2008} or dynamical interactions between a single SMBH or MBHB and a globular cluster which has sunk toward the GC \citep{Capuzzo2015, Fragione2016}. Chief among these alternative mechanisms, however, is the Hills mechanism \citep{Hills1988, Gould2003, Yu2003}, which involves the tidal separation of a stellar binary following a close encounter with a single SMBH. Indeed, it is difficult to contrive a scenario wherein ejections via the MBHB slingshot ejection occur but ejections via the Hills mechanism do not. While there are intrinsic differences in the kinematics and properties of stars ejected via the two mechanisms \citep[see][and references therein]{Brown2015rev, Rasskazov2019}, in practice this is difficult to discern directly, given that current observations can detect only a biased subsample of these populations, i.e. those which are bright and relatively nearby. When considering \textit{null} HVS detections this is not a problem, as a complete dearth of HVSs can constrain all ejection mechanisms simultaneously. For \textit{non-null} detections, however, as may be the case for \textit{Gaia} DR4, this becomes worthy of discussion. Confident constraints on an HVS ejection mechanisms cannot be imposed using a particular HVS candidate without a notion of which mechanism(s) may be responsible ejecting it. The HVS candidate S$^5$-HVS1 is an example of this conundrum, and the next subsection below will be devoted to it. While it is beyond the scope of this work to develop a fully self-consistent ejection model including both the Hills mechanism and the MBHB mechanism simultaneously, in this subsection we consider whether HVS populations ejected via these different mechanisms can be meaningfully disentangled.

\begin{figure*}
    \centering
    \includegraphics[width=2\columnwidth]{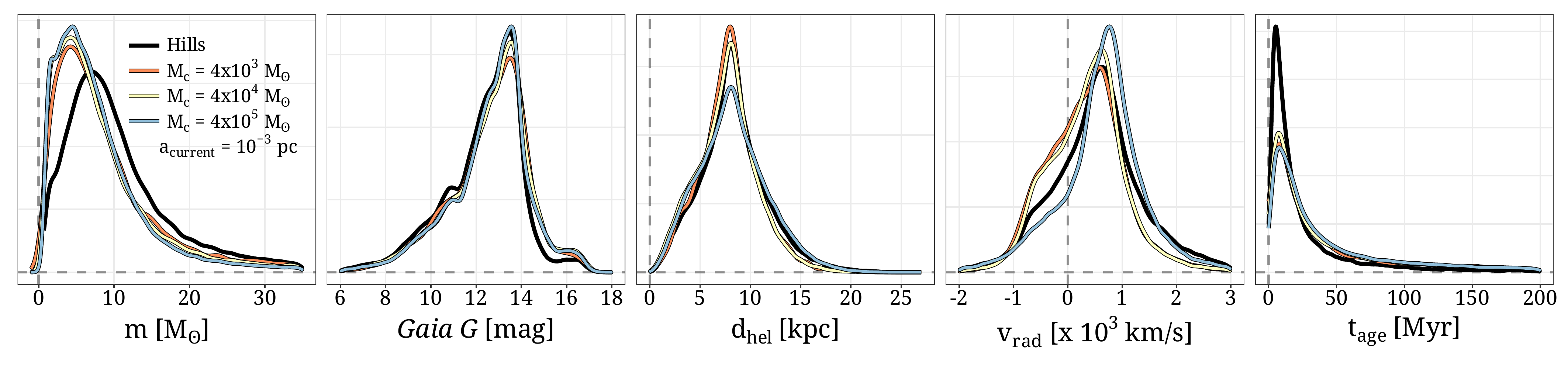}
    \includegraphics[width=2\columnwidth]{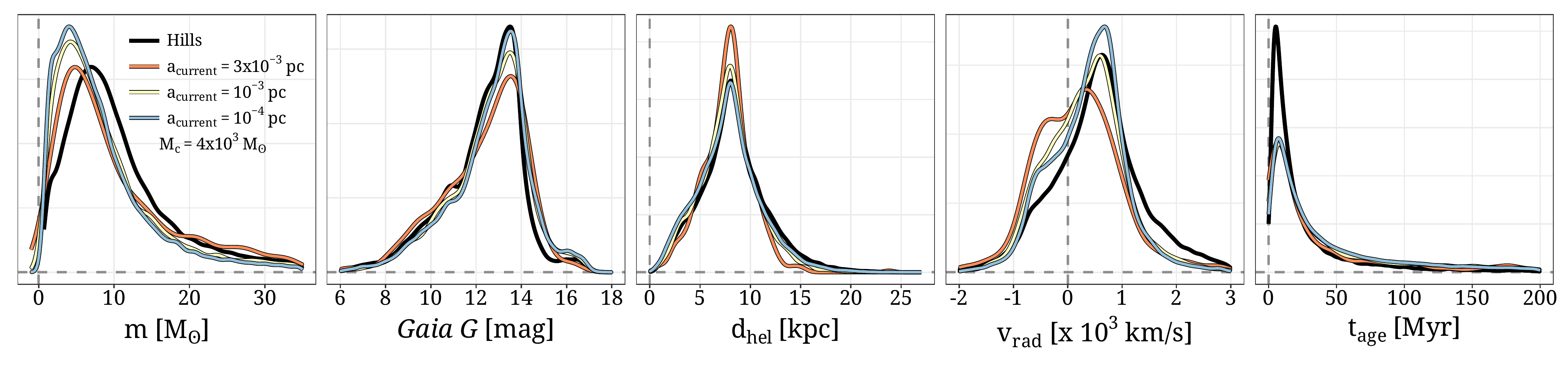}
    \includegraphics[width=2\columnwidth]{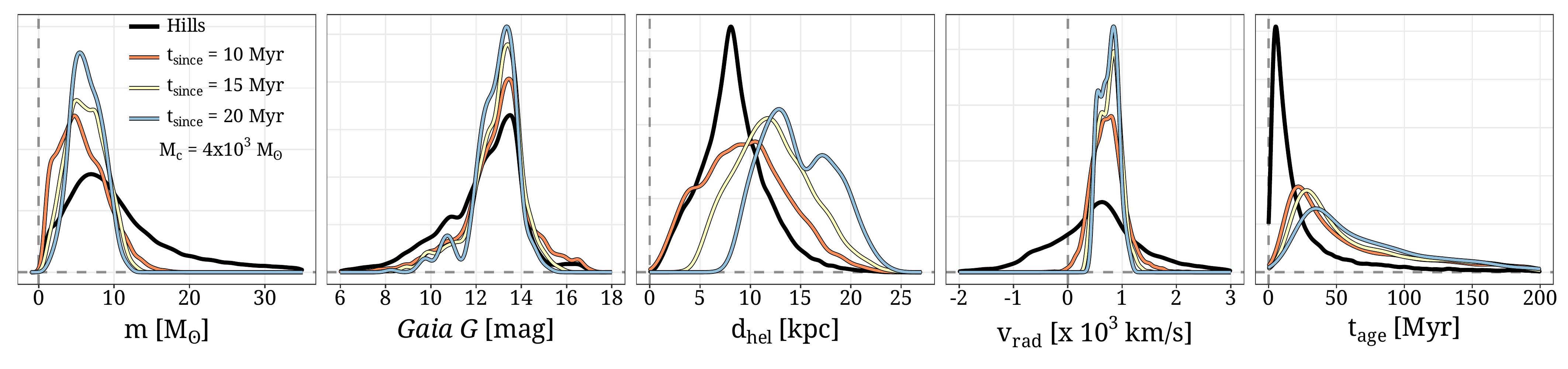}

    \caption{Among HVSs detectable in \textit{Gaia} DR4, distributions of  HVS stellar mass $m$, \textit{Gaia} $G$-band magnitude, heliocentric distance $d_{\rm hel}$, line-of-sight velocity $v_{\rm rad}$ and HVS stellar age $t_{\rm age}$. Black curves show distributions for HVSs ejected via the Hills mechanism and coloured curves show distributions for HVSs ejected via the MBHB slingshot mechanism when the companion mass (top row), MBHB separation (middle row) and time since the MBHB merger (bottom row) are varied. }
    \label{fig:StairStep}
\end{figure*}

To compare the two populations, we use a population of Hills mechanism-ejected HVSs generated by \citet{Marchetti2022}. We refer the reader to that work for more details concerning the generation of this catalogue, but it assumes an IMF index among HVS progenitors of $\kappa=-1.7$ and generates HVS progenitor binaries assuming power-law distributions of the binary mass ratio and log-orbital period. In Fig. \ref{fig:StairStep} we compare the properties of the \textit{Gaia} DR4-detectable HVS population ejected via the Hills mechanism (black) to the population ejected via the MBHB slingshot mechanism (orange/yellow/blue). Shown are the distributions of the detectable stars' stellar masses, \textit{Gaia G}-band magnitudes, heliocentric distances, heliocentric radial velocities and stellar ages, stacked over 40 iterations. When we vary $M_{\rm c}$ or $a_{\rm current}$ (top and middle rows), differences between the populations are subtle. Hills-ejected HVSs are on average slightly more massive and younger than MBHB slingshot-ejected ones, and if the Sgr A* companion is particularly massive and/or its separation from Sgr A* is small, slingshot-ejected HVSs will on average exhibit a narrower range of radial velocities and will span a wider range of distances. 

If Sgr A* and its companion have already merged, differentiating between the mechanisms becomes easier. The bottom row of Fig. \ref{fig:StairStep} shows that as the time $t_{\rm since}$ since coalescence increases, the range of stellar masses among detectable MBHB slingshot-ejected HVSs narrows considerably (massive HVSs will have already left the main sequence, low-mass HVSs will be too far away to detect), the typical distance increases (the final `gasp' of HVSs will be further away), the range of radial velocities narrows (they are less impacted by the Galactic potential and farther away, so their velocity is more in the radial direction) and their typical ages increase (they can't be younger than $t_{\rm since}$). However, given that we expect $5_{-4}^{+11}$ Hills mechanism-ejected HVSs in the radial velocity catalogue of \textit{Gaia} DR4 \citep{Evans2022b} and \textit{at most} a few tens of MBHB slingshot-ejected HVSs (Fig. \ref{fig:NumsDR4}), uncontroversially assigning each HVS to a corresponding ejection mechanism will be difficult in any case.

\subsection{The case of S$^5$-HVS1
}
\label{sec:discussion:S5HVS1}

Thus far in this work, the only HVS observational data we have considered is the lack of high-quality HVS candidates in \textit{Gaia} DR3. Both null \textit{and} non-null HVS detections exist in other searches and surveys, however, which can in principle strengthen constraints. For example, the MMT HVS Survey \citep{Brown2009, Brown2012, Brown2014} identified several dozen HVS candidates. We choose not to include these in our analysis since proper motion measurements for these candidates are not sufficiently precise to conclusively associate any with an ejection from the Galactic Centre\footnote{Note, however, that unless one invokes an interaction with a massive black hole, it becomes quite difficult to explain the extreme velocities of young HVS candidates with well-constrained total velocities such as HVS1 \citep{Brown2005}.}. 

Worth mentioning as well is the HVS candidate S5-HVS1, a $\sim2.35 \, M_\odot$ star first identified by \citet{Koposov2020} in the S$^5$ survey \citep{Li2019}. S$^5$-HVS1 is notable in that it is the first HVS candidate for whom an origin in the GC is uncontroversial -- its trajectory points directly away from the GC and implies an ejection $4.8 \, \mathrm{Myr}$ ago with a velocity of $v_{\rm ej}\simeq 1800 \, \mathrm{km \ s^{-1}}$. As an unambiguous HVS detection, we showed in \citet{Evans2022b} that penalizing models which predict zero or $\gg$1 HVSs similar to S$^5$-HVS1 in the S$^5$ survey significantly improved constraints on the IMF slope in the GC and the ejection rate of HVSs via the Hills mechanism. In this subsection we comment on how and whether the inclusion of this star can improve constraints on a current of former companion to Sgr A*. 

To identify mock HVSs which would have appeared as promising HVSs in S$^5$ by the analysis of \citet{Koposov2020}, we first compute each star's apparent magnitude in the Dark Energy Camera (DECam) $g$ and $r$ bands \citep{DES2018} using the \texttt{MIST} models as done in Sec. \ref{sec:methods:potential}. We then roughly reproduce the S$^5$ selection function, selecting HVSs within the S$^5$ sky footprint \citep[see][table 2]{Li2019} which have \textit{Gaia} parallaxes satisfying $\varpi < 3\sigma_{\varpi} + 0.2$, and DECam photometry satisfying $15<g<19.5$ and $-0.4 < (g-r) < +0.1$. We then take only those mock HVSs with heliocentric radial velocities larger than $800 \, \mathrm{km \ s^{-1}}$, since these were the ones selected for further inspection by \citet{Koposov2020}.

\begin{figure*}
    \centering
    \includegraphics[width=\columnwidth]{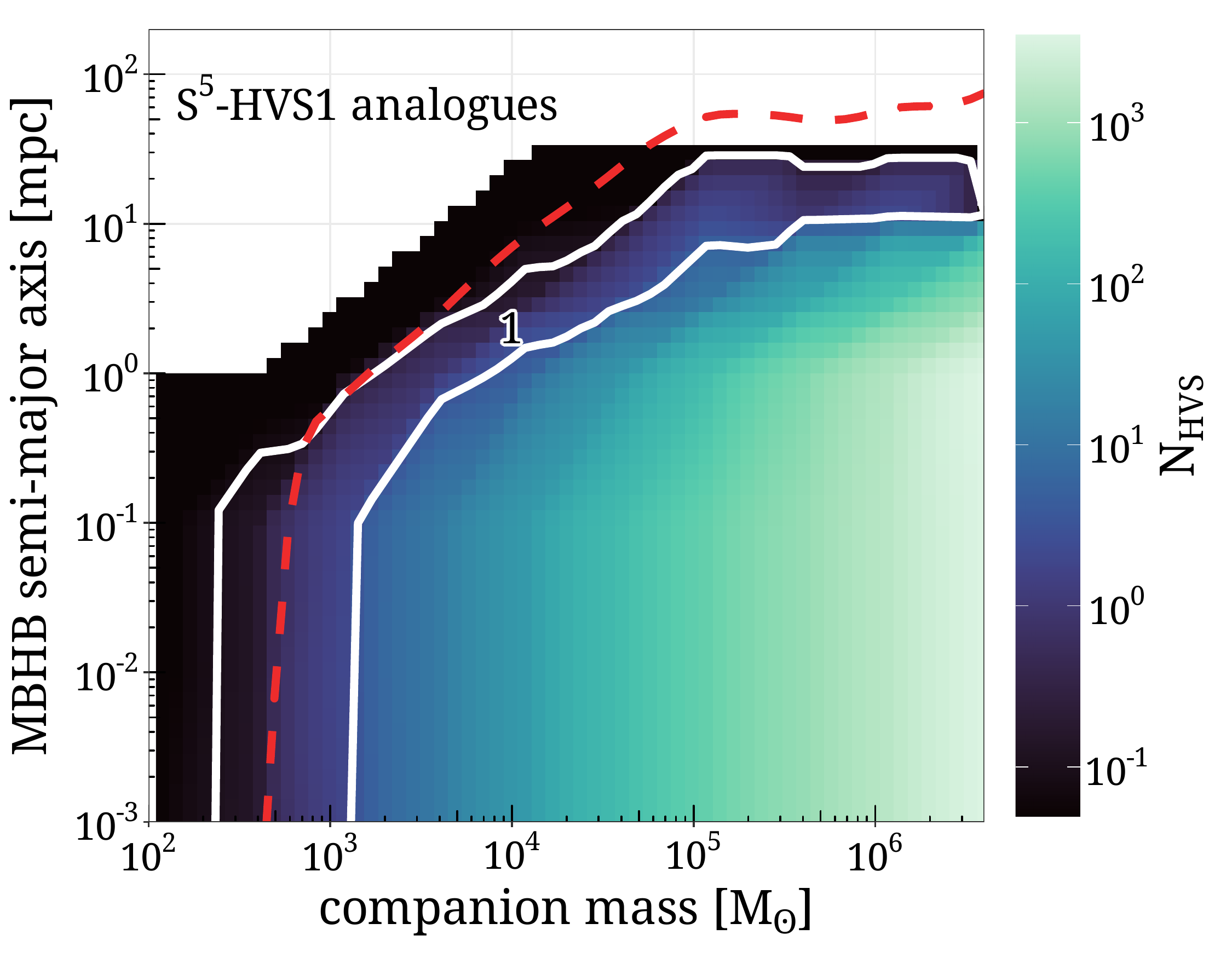}
    \includegraphics[width=\columnwidth]{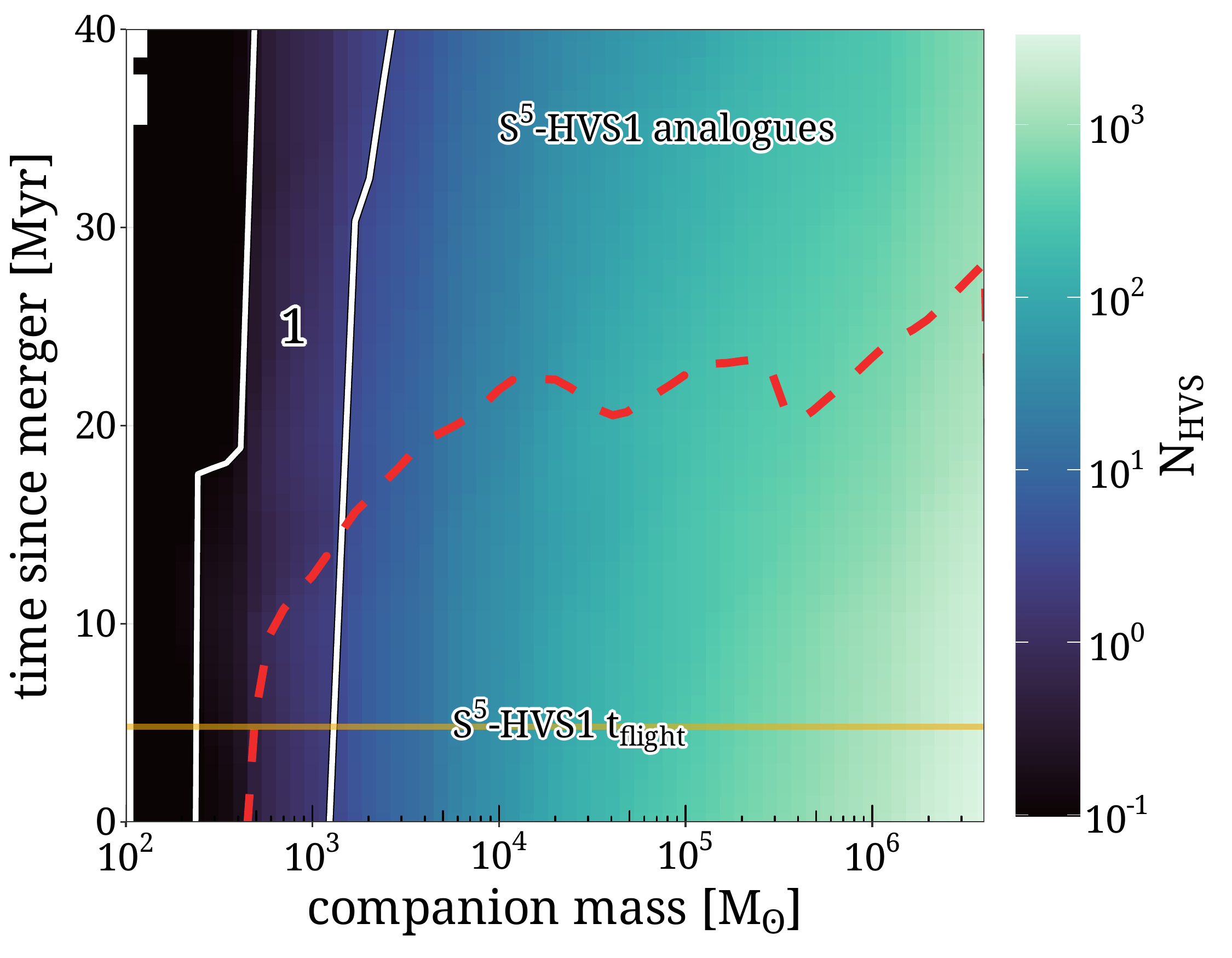}

    \caption{The colorbar shows the population of HVSs predicted to appear in the S$^5$ survey (see Sec. \ref{sec:discussion:S5HVS1} for details) in both $M_{\rm c}$-vs-$a_{\rm current}$ space (left) and $M_{\rm c}$-vs-$t_{\rm since}$ space (right). Configurations below the red dashed lines are excluded by the results of this work. The white, hashed areas show regions of parameter space consistent with finding exactly one HVS in S$^5$. The gold horizontal line in the right plot shows the inferred $S^5$-HVS1 flight time of $4.8 \, \mathrm{Myr}$ \citep{Koposov2020}.}
    \label{fig:S5HVS1}
\end{figure*}

The results of these selections are shown in Fig. \ref{fig:S5HVS1}. Assuming the MBHB slingshot mechanism is the only mechanism ejecting HVSs, the left panel shows the number of S$^5$-HVS1 analogues predicted to appear in S$^5$ depending on the Sgr A* companion mass and the current separation of the MBHB. Configurations between the white lines are consistent within 1$\sigma$ with finding exactly one star similar to S$^5$-HVS1 in S$^5$, and herein lies the issue: a majority of these configurations exist within the parameter space already excluded by the lack of HVSs in the radial velocity catalogue of \textit{Gaia} DR3. Assuming the MBHB slingshot mechanism is the only mechanism responsible for the ejection of HVSs, explaining both the lack of HVSs in \textit{Gaia} DR3 and the existence of S$^{5}$-HVS1 is only possible if Sgr A* has a $200 \, \mathrm{M_\odot} \lesssim M_{\rm c} \lesssim 500 \, \mathrm{M_\odot}$ companion at a separation of $\lesssim 0.3$ pc. 

In the right-hand panel we show similar results in $M_{\rm c} - t_{\rm since}$ space. Sgr A* having merged within the last few ~tens of Myr with a $\sim 200-1000 \, \mathrm{M_\odot}$ companion is consistent with both the existence of S$^5$-HVS1 and the lack of HVSs in DR3. Note, however, that S$^5$-HVS1 has a well-constrained  flight time of 4.8 Myr. If it were ejected via the MBHB slingshot mechanism shortly before a merger, this merger must have occurred at most 4.8 Myr ago. Within this timeframe, only a $200 \, \mathrm{M_\odot} \lesssim M_{\rm c} \lesssim 500 \, \mathrm{M_\odot}$ companion can reconcile the existence of an HVS in S$^5$ with the non-detection of HVSs in \textit{Gaia} DR3.

In conclusion, our simulations only support a MBHB slingshot origin for S$^5$-HVS1 in specific circumstances. Additionally, our criteria to select S$^5$-HVS1 analogues above makes no consideration for its extreme Galactocentric velocity of $1750 \, \mathrm{km \ s^{-1}}$. A velocity this large is quite hard to achieve in the MBHB slingshot mechanism for a companion mass of $\lesssim 500 \, \, \mathrm{M_\odot}$ -- only $\sim$20 per cent of our S$^5$-HVS1 analogues in this mass range are as fast as S$^5$-HVS1. Typical ejection velocities via the Hills mechanism, on the other hand, are comparatively larger and can more easily accommodate this star. A more rigourous investigation of whether S$^5$-HVS1 could be produced by this mechanism is warranted and does not yet exist. \citet{Generozov2020} used S$^5$-HVS1's flight time as an observational constraint and found that a $1000 \, \mathrm{M_\odot}$ companion separated from Sgr A* by 0.01 pc could efficiently reproduce the observed eccentricity distribution of the S-star cluster in the GC. This, however, does not necessarily imply S$^5$-HVS1 was ejected via the MBHB slingshot mechanism.

\subsection{The impact of model simplifications}

Our model for the ejection of HVSs from the MBHB slingshot mechanism includes a number of implicit and explicit simplifying assumptions. Works in recent years have shown that some of these assumptions i) impact the evolution of MBHBs, and ii) may not hold strictly true in the centre of our own Galaxy. In this subsection we comment on the degree to which our results could be impacted by these simplifications.


We assume that our MBHB is non-accreting and in a gas-free environment. Dynamical friction driven by a dense gaseous disc and accretion onto the MBHB impact the separation, eccentricity and inclination evolution of the MBHB \citep[see][]{Escala2005, Dotti2006, Dotti2007, Munoz2019} and would influence the stellar evolution of HVS progenitors \citep[see][]{Cantiello2021}. Accretion onto the MBHB in the GC would not be strong enough to impact its evolution -- X-ray observations of the Galactic Centre indicate an accretion rate of $10^{-6} - 10^{-5} \, \mathrm{M_\odot \ yr^{-1}}$ \citep{Baganoff2003, Quataert2004} at the Bondi radius ($0.04 \, \mathrm{pc}$) and polarization measurements limit the accretion rate to $\sim10^{-8} \, \mathrm{M_\odot \ yr^{-1}}$ near the Schwarzschild radius \citep{Quataert2000, Bower2003, Marrone2007}. We can check the impact of this accretion with a back of the envelope calculation. Assuming that the inspiral of a MBHB is driven entirely by gas accretion, differentiating the angular momentum of the MBHB with respect to time yields
\begin{equation}
    \frac{\dot{a}}{a} = 2\left(\frac{\dot{\ell}}{\ell} - \frac{3}{2}\right) \frac{\dot{M}}{M} \; \text{,} 
\end{equation}
where $\dot{a}$ is the hardening rate, $\dot{M}$ is the rate of accretion onto the MBHB, $M$ is the total MBHB mass and $\ell = L/M$ is the specific angular momentum of the binary. $2(\dot{\ell}/\ell - 3/2)$ is a constant of order unity depending on gas accretion physics \citep[see][]{DOrazio2021}. Plugging in characteristic numbers for the accretion rate and MBHB mass, the accretion-driven hardening rate is never more than ten orders of magnitude smaller than the total slingshot + gravitational wave hardening rate. While evidence exists for an episode of nuclear activity in the GC in the last few Myr \citep[see][and references therein]{BlandHawthorn2019}, this flare was short-lived.  On the scale of detectable HVS flight times, it is valid to assume a putative MBHB in the GC would be non-accreting and in a gas-poor environment. See \citet{Naoz2020} for further discussion on the impact of accretion on a presumed black hole companion to Sgr A*.

We have assumed for simplicity that the presumed MBHB in the GC is on an initially circular orbit and never deviate from a circular orbit until coalescence. This assumption is valid since while non-zero eccentricity increases the binary hardening rate at small separations, it leads to only minor variations in the total stellar mass ejected via the slingshot mechanism \citep{Quinlan1996, Sesana2006, Rasskazov2019}. In any event, the impact of eccentricity is only relevant for MBHB mass ratios larger than 10$^{-3}$ -- smaller mass ratio MBHBs tend to circularize with time \citep{Rasskazov2019, Bonetti2020}.

Our model of the GC assumes the MBHB is embedded within a spherical, nonrotating nuclear star cluster (NSC). In actuality, the NSC is slightly flattened along the Galactic vertical axis \citep{Schodel2014} and rotates more or less parallel with the Galactic disc \citep{Trippe2008, Schodel2009, Chatzopoulos2015, Fritz2016}. A nuclear cluster which co-(counter-)rotates with the MBHB enhances (suppresses) eccentricity growth of the MBHB \citep{Sesana2011, Rasskazov2017, Rasskazov2019, Bonetti2020} while the impact of counter- and co-rotation of the NSC on the MBHB hardening rate is less clear \citep[see contradictory results in][]{HolleyBokelmann2015, Mirza2017, Rasskazov2017, Rasskazov2019, Varisco2021}. The MBHB slingshot mechanism mass ejection rate is decreased slightly at small separations ($a\lesssim 0.3 a_{\rm h}$) if the NSC is maximally corotating \citep{Rasskazov2019}, in which case our assumption of a non-rotating NSC may be overestimating $N_{\rm HVS}$. If the NSC is maximally counter-rotating, the mass ejection rate at small separations is increased by a factor of up to $\sim$two \citep{Rasskazov2019}. This is less of a concern, given that MBHBs are expected to align their rotation with the angular momentum of the NSC (see below). 

The density $\rho$ and velocity dispersion $\sigma$ of the NSC impact the MBHB hardening rate (Eq. \ref{eq:dadt_HVS} - Eq. \ref{eq:ah}); we assume single values for these, when in fact they both vary with radial distance from the GC \citep{Schodel2009, Schodel2014, Feldmeier2014}. We have confirmed that variations of $\sigma$ and $\rho$ within observational uncertainties do not meaningfully affect the results of this work. Along with this, we neglect to consider mass segregation as well within the NSC. A mass segregated stellar environment accelerates hardening timescales among highly unequal-mass MBHBs but has the opposite impact on $\sim$equal mass MBHBs \citep{Mukherjee2023}. We assume as well that the MBHB centre of mass is always centred on the exact geometric centre of the Galaxy. In reality, the MBHB will `drift' randomly in a nonrotating stellar nucleus \citep{Merritt2001, Chatterjee2003} or on a closed orbit if the nucleus is rotating \citep{HolleyBokelmann2015, Mirza2017, Khan2020, Varisco2021}. The center of mass displacement from the Galactic barycentre is typically on the order of the MBHB radius of influence or smaller, i.e. not large enough to meaningfully affect the detectable HVS population. There are very few cases where shifting the positions of a \textit{Gaia}-detectable mock HVS by $\sim$a few pc in any direction renders it undetectable.

When generating mock HVS populations, we take no consideration of the orientation and phase of the MBHB. If stars are ejected isotropically in the MBHB slingshot mechanism, the orientation of the MBHB is irrelevant. While our model assumes isotropic ejection, theoretically the fastest-ejected stars are ejected preferentially in the plane of the binary, with a complex polar angle distribution depending on its separation, eccentricity and mass ratio \citep{Sesana2006, Rasskazov2019, Darbha2019}. Ejections are likely non-axisymmetric as well, though the azimuthal distribution of fast ejections remains unclear \citep[see discussion in][]{Rasskazov2019}. Without independent constraints on the inclination/phase of the MBHB, we would be compelled to sample them at random, and therefore any angular dependence would get washed out in our results after averaging over many iterations. There is reason to believe, however, that a putative MBHB in the GC would be aligned with the Galactic disc. For a spherical, nonrotating NSC, the orientation of the MBHB orbit drifts on the order of $\sqrt{m_{*}/M_{\rm MBHB}}\sim10^{-4}$ rad relative to its initial orientation, where $m_{*}$ is the typical mass of a star in the NSC and $M_{\rm MBHB}$ is the total mass of the MBHB \citep{Merritt2001}. In a rotating NSC, however, the angular momentum of the MBHB aligns with the angular momentum of the nucleus \citep{Gualandris2012, Wang2014, Cui2014, Rasskazov2017}. Having established above that the Milky Way NSC rotates parallel to the Galactic disc, it is not unreasonable to assume a putative MBHB would rotate parallel with the disc as well. Another point in favour of this is the existence of the nuclear stellar disc embedded within the NSC itself \citep{Bartko2009, Lu2009, Yelda2014}. This stellar disc rotates in the direction of the overall Galactic disc as well \citep{Schonrich2015, Schultheis2021, Sormani2022} and resonant relaxation processes would align any IMBH-SMBH binary with such a disc \citep{Szolgyen2021, Magnan2022}. If the MBHB orbit were indeed aligned with the Galactic midplane, our assumption of isotropic ejections would slightly underestimate $N_{\rm HVS}$, since in actuality a greater proportion of HVSs are ejected on angles pointed more flattened toward the Earth.

Finally, our model assumes the phase space of low-angular momentum orbits which bring GC stars on a close ($\lesssim a$) approach to the MBHB is efficiently repopulated over time. Only stars within this `loss cone' experience strong dynamical interaction with the MBHB. This loss cone is refilled over time, primarily via two-body relaxation in the NSC Sgr A* \citep[see][]{Lightman1977, Merritt2013} or resonant relaxation processes in the young clockwise disc \citep{Rauch1996, Madigan2009, Madigan2011, Madigan2014}. In principle, if these scattering mechanisms are not efficient, the loss cone can empty and the MBHB hardening (and therefore the ejections of HVSs) can slow down or cease entirely \citep{Milosavljevic2003}. This problem was investigated in the context of HVS ejections by \citet{Sesana2007losscone}, who found that without loss cone refilling the MBHB stalls after ejecting only half the MBHB reduced mass in stars, $\sim 10^2 - 10^6 \, \mathrm{M_\odot}$ in our case depending on the assumed mass ratio. This is smaller than the total mass ejected in the full loss cone regime by a factor of a few ($\sim$10) if the mass ratio is small (large) and the ejected stars would have smaller typical velocities. Concerns about loss cone depletion among MBHBs has largely been alleviated in recent years, as simulations have shown that relaxation-driven refilling of the loss cone is sufficient to avoid loss cone depletion in triaxial potentials \citep{Berczik2006, Vasiliev2015, Gualandris2017}, or even biaxial potentials \citep{Khan2013}. Regardless, our models in this work assume a full loss cone, and this can lead to an overestimation of $N_{\rm HVS}$ in cases of partial depletion. This assumption is likely the largest source of systematic uncertainty in our modelling. Given the strong dependence of $N_{\rm HVS}$ large range of HVS population sizes  As noted by \citet{Vasiliev2015}, it is relatively straightforward to adjust $H$ and $J$ (Eq. \ref{eq:HJfits}) to account for loss cone depletion \citep[see also][]{Rasskazov2019}. 
\section{Conclusions}
\label{sec:conclusions}

Massive black hole binaries (MBHBs) are a natural consequence of galaxy evolution. Dynamical interactions between stars and an MBHB in the centre of the Milky Way could eject hypervelocity stars (HVSs) detectable by the \textit{Gaia} space satellite. In this work, we use existing HVS observations for the first time as a probe of a possible supermassive or intermediate mass companion black hole to Sgr A*, the supermassive black hole located in the Galactic Centre (GC). Building upon previous work, we realistically simulate the ejection of HVSs from an MBHB assuming a variety of MBHB mass ratios and separations. We focus in particular on HVSs which would have appeared in the radial velocity catalogue of the third data release from \textit{Gaia}. Considering that zero HVSs with precise astrometry were unearthed in the radial velocity catalogue of \textit{Gaia} DR3 \citep{Marchetti2022}, MBHB configurations which predict too many HVSs in this data release can be excluded. Our conclusions are as follows:
\begin{itemize}
	\item The number of HVSs detectable by \textit{Gaia} depends strongly on the MBHB separation and companion mass. It is comparatively less sensitive to the shape of the assumed initial mass function (IMF)  (Fig. \ref{fig:Nums})
        \item For a fiducial Sgr A* companion mass of $4000 \, \mathrm{M_\odot}$ and MBHB separation of $0.001 \, \mathrm{pc}$, $8 \pm 4$ HVSs should have been detected in the \textit{Gaia} DR3 radial velocity catalogue with precise astrometry (Fig. \ref{fig:McaDR3}). 
        \item The \textit{lack} of such HVSs in DR3 excludes a companion within $1 \, \mathrm{mpc}$ of Sgr A* unless it has a mass of $\lesssim 1000 \, \mathrm{M_\odot}$, complementing and extending prior constraints on a Sgr A* companion (Fig. \ref{fig:CompanionConstraints}).
        \item The lack of confident HVS detections in \textit{Gaia} DR3 also allows us to constrain a \textit{former} companion to Sgr A* for the first time. If Sgr A* merged with a companion in the recent past, either of the following must be true: i) the former companion had a mass of $\lesssim500 \, \mathrm{M_\odot}$, or ii) the merger must have happened more than $\sim$10-30 Myr ago (Fig. \ref{fig:MctDR3}).
        \item If Sgr A* has an existing companion or had a former companion, the forthcoming fourth \textit{Gaia} data release will contain at most a few tens of HVSs ejected via the MBHB slingshot mechanism in its radial velocity catalogue with precise astrometry (Fig. \ref{fig:NumsDR4}).
\end{itemize} 

The constraints we place in this work on a possible companion to Sgr A*, especially when combined with constraints obtained over the last few years \citep{Naoz2020, Reid2020, GRAVITY2020}, appear to be closing the door on the existence of such a companion. While a massive companion up to $\sim10^5 \mathrm{M_\odot}$ is still allowed at separations larger than $\sim0.1 \, \mathrm{pc}$, a hardened MBHB in the GC appears unlikely unless the MBHB mass ratio is extreme. With the future \textit{Gaia} data releases and their synergy with both forthcoming and currently operational spectroscopic facilities and surveys  (e.g. WEAVE; \citealt{Dalton2012}, 4MOST; \citealt{deJong2019}, SDSS-V MWM; \citealt{Kollmeier2017}), more and more HVSs will be detected. If Sgr A* does not have a companion of significant mass and did not have one in the recent past, the MBHB slingshot mechanism can be definitely ruled out as an avenue for HVS ejections and future research can focus on more realistic mechanisms.

\section*{Acknowledgements}

The authors thank J.R. Westernacher-Schneider for helpful discussion. FAE acknowledges support from the University of Toronto Arts \& Science Postdoctoral Fellowship program and the Dunlap Institute. TM acknowledges an ESO fellowship. EMR acknowledges that this project has received funding from the European Research Council (ERC) under the European Union’s Horizon 2020 research and innovation programme (grant agreement No. 101002511 - VEGA P). JB acknowledges financial support from NSERC (funding reference number RGPIN-2020-04712).

\section*{Data Availability}

The simulation outputs underpinning this work can be shared upon reasonable request to the corresponding author. \textcolor{black}{These simulations were produced using the \texttt{speedystar} package, publicly available at \url{https://github.com/fraserevans/speedystar}}.


\bibliographystyle{mnras}
\bibliography{HRS}

\bsp	
\label{lastpage}
\end{document}